\documentclass[structabstract]{aa}
\usepackage{txfonts}
\usepackage{graphicx}
\usepackage{natbib}
\bibpunct{(}{)}{;}{a}{}{,}
\title{ Nitrogen chronology of massive main sequence stars}
\author{K. K\"ohler\inst{1}
\and M. Borzyszkowski\inst{1}
\and I. Brott\inst{2}
\and N. Langer\inst{1}
\and A. de Koter\inst{3,4}}
 
\institute{Argelander-Institut f\"ur Astronomie der Universit\"at Bonn, Auf dem H\"ugel 71, 53121 Bonn, Germany
\and University of Vienna, Department of Astronomy, T\"urkenschanzstr.17, 1180 Vienna, Austria
\and Astronomical Institute, Utrecht University, Princetonplein 5, 3584CC, Utrecht, The Netherlands
\and Astronomical Institute Anton Pannekoek, University of Amsterdam, Kruislaan 403, 1098 SJ, Amsterdam, The Netherlands
}
 
\date{Received date / Accepted date}
 
\abstract {Rotational mixing in massive main sequence stars is predicted to monotonically increase their surface nitrogen 
abundance with time.} 
{We use this effect to design a method for constraining the age and the inclination angle of massive main sequence stars, 
given their observed luminosity, effective temperature, projected rotational velocity and surface nitrogen abundance.}
{This method relies on stellar evolution models for different metallicities, masses and rotation rates.
We use the population synthesis code STARMAKER to show the range of applicability of our method.}
{We apply this method to 79 early B-type main sequence stars near the LMC clusters NGC 2004 and N 11 and the SMC 
clusters NGC 330 and NGC 346. From all stars within the sample, 17 were found to be suitable for an age analysis. For ten 
of them, which are rapidly rotating stars without a strong nitrogen enhancement, it has been previously concluded that 
they did not evolve as rotationally mixed single stars. This is confirmed by our analysis, which flags the age of these 
objects as highly discrepant with their isochrone ages. For the other seven stars, their nitrogen and isochrone ages are 
found to agree within error bars, what validates our method. Constraints on the inclination angle have been derived for 
the other 62 stars,with the implication that the nitrogen abundances of the SMC stars, for which mostly only upper
limits are known, fall on average significantly below those limits.}
{Nitrogen chronology is found to be a new useful tool for testing stellar evolution
and to constrain fundamental properties of massive main sequence stars. A web version of this tool
is provided.} 
 
 
\begin{document}
\maketitle
 
\titlerunning{Nitrogen chronology of massive MS stars}
 
\authorrunning{K. K\"ohler et al.}
 
\section{Introduction}
\label{sec1}
The age determination of stars is an important objective, and several methods are currently being used. 
For star clusters, which presumably contain stars of similar ages, comparing the main sequence turn-off
point in a color-magnitude or Hertzsprung-Russell diagram with stellar evolution models can predict
the age of all stars in the cluster \citep{Jia_2002,Buonanno_1986}. 
For individual stars, ages can be derived by comparing their location in the HR-diagram with isochrones computed
from stellar evolution models. In the envelope of low mass main sequence stars lithium is depleted, 
resulting in a monotonous decrease of 
the surface lithium abundance with time, which can be used to determine their age \citep{Soderblom_2010,Randich_2009}. 
 
Massive stars fuse hydrogen to helium via the CNO cycle. The reaction 
$\element[][14]{N} \left(\mathrm{p},\gamma\right) \element[][15]{O}$ has the lowest cross section, resulting in an increase of the 
nitrogen abundance compared to carbon. The evolution of massive stars is 
furthermore thought to be affected by rotation, which may cause mixing processes including large scale circulations and local 
hydrodynamic instabilities \citep{Endal_1987}. The nitrogen produced in the core can be 
transported to the surface when rotational mixing is considered, causing a monotonic increase of the surface nitrogen 
abundance with time \citep{Heger_2000,Maeder_2000a,Przybilla_2010}. Here, we suggest to utilize this effect to constrain 
the ages of stars. 
 
Based on detailed stellar evolution models, we present a method which reproduces the surface nitrogen abundances of 
massive main sequence stars as a function 
of their age, mass, surface rotational velocity and metallicity. Three stellar evolution grids for the 
Milky Way (MW), the Large Magellanic Cloud (LMC) and the Small Magellanic Cloud (SMC) are used, providing stellar 
models of different metallicities, masses and rotational velocities and include the effects of rotational mixing 
\citep{Brott_2010a}. The mixing processes in these stellar evolution models were calibrated to reproduce 
the massive main sequence stars 
observed within the VLT-FLAMES Survey of Massive Stars \citep{Evans_2005,Evans_2006}.
 
The abundance analysis of massive stars is often focused on stars with small $v \sin i$ values,
as this allows to obtain abundances with higher accuracies \citep{Przybilla_2010}.
However, those objects are not well suited for nitrogen chronology, as they are either slow rotators,
or have a low inclination angle. In the latter case, the age can not be well constrained, since a
large range of true rotational velocities is possible. 
Many stars analyzed in the VLT-FLAMES Survey, however, are particularly suited as targets for nitrogen chronology, 
since this survey did not concentrate on the apparent slow rotators.

On the other hand, results from the VLT-FLAMES Survey identified two groups of early B-type stars
which could not be reconciled by the 
evolutionary models of single stars with rotational mixing \citep{Hunter_2008,Brott_2010b}.
The first group consists of slowly rotating stars which are nitrogen rich, to which the method of nitrogen
chronology is not applicable. The second group contains relatively fast rotating stars with only a modest 
surface nitrogen enhancement. While such stars are in principle predicted by models of rotating stars
\citep{Brott_2010a}, they were not expected in be found in the VLT-FLAMES Survey due to the 
imposed V-band magnitude cut-off, since unevolved main sequence stars are visually faint, and the visually 
brighter more evolved rapid rotators are expected to be strongly nitrogen enriched \citep{Brott_2010b}.
\citet{Brott_2010b} suggested that either these stars are the product of close binary evolution, or that
rotational mixing is much less efficient than predicted by the stellar evolution models. 

The nitrogen chronology is able to shed new light on this issue. In contrast to the 
population synthesis pursued by \citep{Brott_2010b}, it does not use statistical arguments 
but is applied to each star individually. Its results can therefore be easily compared
with classical age analyses, e.g., using isochrone fitting in the HR diagram.
In this paper, we apply the nitrogen chronology to all LMC and SMC stars from the VLT-FLAMES Survey which have
a projected rotational velocity larger than 100\,km/s. 
 
We organize our paper as follows.
In Sect.~\ref{sec2} our method of nitrogen chronology is presented. 
Sect.~\ref{sec3} describes our analysis of the VLT-FLAMES sample. A discussion and evaluation 
of our new method is given in Sect.~\ref{sec4} and \ref{sec5}. Sect.~\ref{sec6} will round off with our 
conclusions.

\section{Method}
\label{sec2}
\begin{figure*}[htb]
    \centering
      \includegraphics[width=17cm]{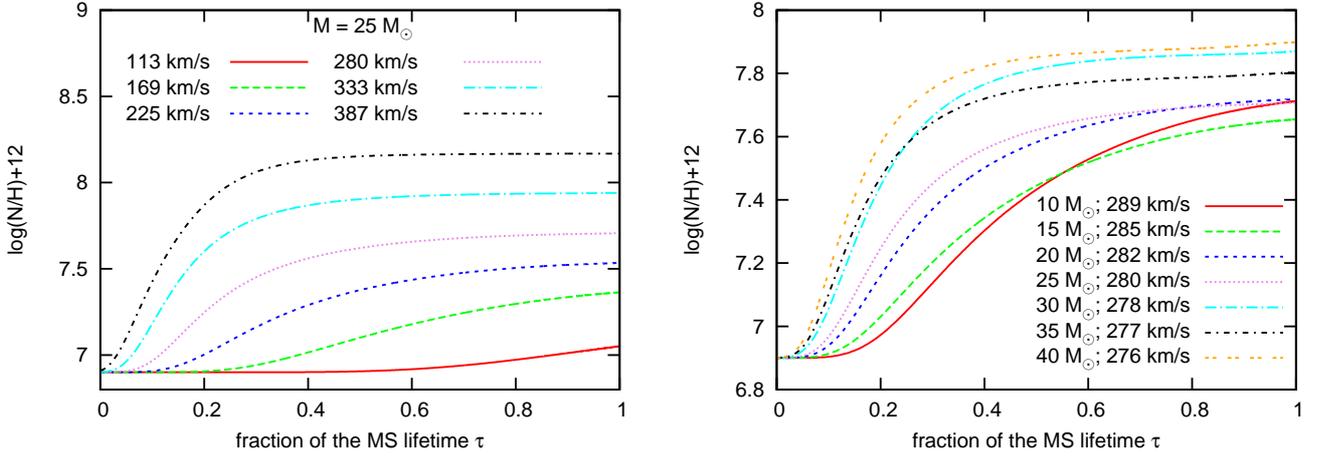}
    \caption{For different models of the stellar evolution grid of \citet{Brott_2010a} using LMC metallicity, the surface nitrogen 
abundance is plotted against the fractional main sequence lifetime. The left panel shows the influence 
of the initial rotational velocity $v_\mathrm{ZAMS}$ for a given mass of $M=25\,M_{\odot}$. The right panel 
depicts the evolution of the surface nitrogen abundance for several masses at $v_\mathrm{ZAMS} \approx 280\,$km/s.}
    \label{fig_nitrogen}
\end{figure*}

The enrichment of nitrogen at the surface of a massive star during its evolution on the main sequence depends on its 
mass and rotation rate. In Fig.~\ref{fig_nitrogen} the surface nitrogen abundance is plotted against the fractional 
main sequence lifetime $\tau = t/\tau_\mathrm{H}$  for different initial masses and surface rotational velocities.
Here, $\tau_\mathrm{H}$ is the hydrogen burning lifetime of a stellar evolution sequence. A parameterization of $\tau_\mathrm{H}$ as 
a function of mass and surface rotational velocity, for all three metallicities considered here, is given in 
Appendix~\ref{sec2_Cmslairv}.
For our method, we define ``zero-age'' as the time when the mass fraction of hydrogen has decreased by 2\% in the
center due to hydrogen core burning. This ensures that the stellar models are still unevolved but already thermally relaxed.
The surface nitrogen abundance $\mathcal N$ is given by 
\begin{equation}
\mathcal N = \log \left( \frac{N}{H} \right) + 12,
\label{eq_SNA}
\end{equation}
with the surface number density of nitrogen and hydrogen atoms $N$ and $H$, respectively. 
 
Figure~\ref{fig_nitrogen} shows that the nitrogen abundance rises smoothly over time starting with the surface nitrogen 
abundance at the ZAMS depending on the metallicity. It reaches its maximum value at the end of the main sequence 
evolution. The right panel of Fig.~\ref{fig_nitrogen} depicts the mass dependence of the nitrogen enrichment at the surface. 
For a given initial surface rotational velocity, the models show a faster enrichment at the surface for higher initial 
mass. Furthermore, the faster the initial surface rotational velocity for a given mass the greater is the effect of rotational 
mixing which leads to faster enrichment and a higher saturation value \citep{Heger_2000,Meynet_2000,Maeder_2001}.

To constrain the age of an observed massive main sequence star using its surface nitrogen abundance, we derived an equation 
which fits the surface nitrogen abundance for a given mass, surface rotational velocity and metallicity to the models 
of rotationally mixed massive stars. In order to reduce interpolation errors in constructing this equation, stellar evolution grids
are required which have a dense spacing in the grid parameters initial mass $M$ and initial equatorial rotational
velocity $v_\mathrm{ZAMS}$. Presently, stellar evolution grids available for this purpose are those
presented in \citet{Brott_2010a}.

The surface nitrogen abundance monotonically increases over the fraction of the main sequence lifetime 
(cf. Fig.~\ref{fig_nitrogen}). This can be reproduced well with an ansatz as
\begin{equation}
\mathcal N = a + c \cdot \gamma_d \left( b \cdot \tau \right).
\label{eq_SNA_1}
\end{equation}
The parameters $a$, $b$, $c$, and $d$ are functions of the initial surface rotational velocity $v_\mathrm{ZAMS}$ and the 
initial mass $M$ of a star. The incomplete gamma function $ \gamma_d \left( b \cdot \tau \right)$ depends on the product 
$ b \cdot \tau $, with the fraction of the main sequence lifetime $\tau$ and the parameters $b$ and $d$. For calculating 
the surface nitrogen abundance according to Eq.~(\ref{eq_SNA_1}) the incomplete gamma function is normalized by the 
complete gamma function $\Gamma$ as
\begin{equation}
\gamma_d \left( b \cdot \tau \right) = \frac{1}{\Gamma \left( d \right)} \int_0^{b \cdot \tau} x ^{d-1} e^{-x} \,\mbox{d}x.
\label{eq_SNA_2}
\end{equation}

Considering the properties of the incomplete gamma function it is possible to make some statements about the parameters 
$a$, $b$, $c$, and $d$. The values of $ \gamma_d $ range from 0 to 1, so $c$ is responsible for the height of the curve. 
The parameter $a$ describes the initial surface nitrogen abundance, which is independent of mass and surface rotational 
velocity. The parameter $b$ describes the speed of the nitrogen enrichment i.e. the steepness of the
curves in Fig.~\ref{fig_nitrogen}.

The parameters $b$ and $c$ depend on initial mass and rotational velocity. By fitting Eq.~(\ref{eq_SNA_1}) to the 
data of the stellar evolution models, both parameters can be determined for different values of $M$ and $v_\mathrm{ZAMS}$.
Considering the parameter $d$, the best fits of Eq.~(\ref{eq_SNA_1}) to the data of the three stellar 
evolution grids are obtained by a linear dependence of $d$ on the surface rotational velocity $v_\mathrm{ZAMS}$ 
(see Appendix~\ref{app1}).

Nitrogen profiles calculated by using Eq.~(\ref{eq_SNA_1}) can be useful for binary population studies, where 
a quick knowledge of the surface nitrogen abundance is required.

\subsection{Accuracy}
\begin{table}[h]
\caption{Parameters of the example sequence discussed in the text and displayed in Fig.~\ref{fig_comp}.}
\label{tab_compare}  
\centering                                  
\begin{tabular}{l r r}  
\hline\hline
parameter & Model 2 & Model 1 \\
\hline
$M$ [$M_{\odot}$] & 30 & 35 \\
$v_\mathrm{ZAMS}$ [km/s] & 169 & 168 \\
$Z$ & $Z_{\mathrm{LMC}}$ & $Z_{\mathrm{LMC}}$\\
$a$ & 6.9 & 6.9 \\
$b$ & 12.8 & 14.6 \\
$c$ & 0.6 & 0.6 \\
$d$ & 6.0 & 6.0 \\
\hline
\end{tabular}
\end{table}

\begin{figure*}[htb]
    \centering
        \includegraphics[width=17cm]{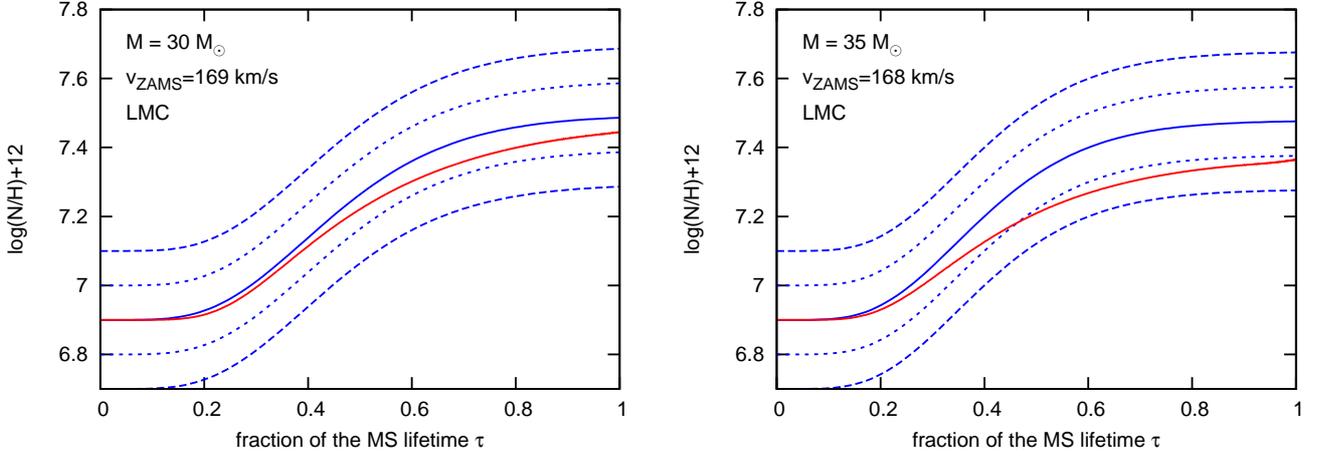}
    \caption{Surface nitrogen abundance as function of the fractional main sequence lifetime 
for two specified stellar models (see legend) from the detailed stellar evolution grids
of \citet{Brott_2010a} (red lines), compared to the results of our fit (blue solid lines). 
The blue dashed lines represent error ranges of $\pm 0.1$ and $\pm 0.2$~dex for the nitrogen abundance.
While the left panel shows a typical fit, the right panel depicts our worst case.}
    \label{fig_comp}
\end{figure*}

Two stellar models are chosen to indicate the quality of the presented method to calculate the surface nitrogen abundance.
All parameters of Eq.~(\ref{eq_SNA_1}) are calculated from
Eq.~(\ref{eq_paraa_comp} - \ref{eq_parad_comp}) given in Appendix~\ref{app1} (Table~\ref{tab_compare}).
Using these values, the surface nitrogen abundance of the two sequences as a function of $\tau$ 
is calculated and shown as blue solid lines in Fig.~\ref{fig_comp}. 
It is compared to the nitrogen abundance evolution which results from the detailed
stellar evolution models of \citet{Brott_2010a}.

All calculated surface nitrogen abundance functions fit the detailed models very well for low values of $\tau$, 
but show different levels of deviations towards the end of the main sequence evolution. In Fig.~\ref{fig_comp}, a 
typical fit is shown on the left panel. At all times, the fit result deviates by less than 0.1~dex from the detailed model
prediction. The right panel of 
Fig.~\ref{fig_comp} shows our worst case. Even in this case, the fitted surface nitrogen abundance
deviates by more than about 0.1\,dex from the value of the detailed stellar evolution model for a fraction of the time.

Numerical noise in the detailed stellar evolution models does in rare cases lead to a non-monotonic behaviour
of the nitrogen evolution as function of the initial rotation rate or as function of mass (cf., left panel of
Fig.~1). These features are smoothed out by our analytic fitting method, which guarantees monotonic behaviour
in the considered parameter space (Sect. 2.2). Therefore, a deviation of our nitrogen prediction from that of the detailed
stellar evolution models does not necessarily flag a problem of our method. 
In summary, we consider the error in the predicted nitrogen abundance
introduced by our fitting method to be smaller than 0.1~dex.
In the present paper, the surface nitrogen abundances of the analyzed stars have an observational error
of $\pm 0.2\,$dex \citep{Hunter_2007}.

\subsection{Validity range}
\label{valid}
As the lowest mass of the underlying stellar evolution models used in our fitting method is 5\,$M_{\odot}$, this 
defines the lower mass limit for its applicability. Towards very high masses, effects of stellar wind mass loss 
produce features which are not captured by our method above an upper limiting mass, which depends on metallicity 
(see Table~\ref{tab_valcond}).

The stellar evolution models of \citet{Brott_2010a} for very rapidly rotating stars may undergo so called 
quasi-chemically homogeneous evolution. The behaviour of those models is also not captured by our fitting method. 
This problem is avoided when only initial rotational velocities below 350\,km/s are considered, which thus 
defines the upper velocity limit for our method. This also ensures that 2D effects of very rapid rotation on the 
stellar atmosphere do not play a role here.

\begin{table}[h]
\caption{Range of initial stellar parameters for which our method is validated, for the 
three considered metallicities. In addition, for stars more massive than 30\,$M_\odot$, the effective temperature 
should be higher than 25\,000\,K.} 
\label{tab_valcond}  
\centering                                  
\begin{tabular}{l c c}  
\hline\hline
Galaxy & $M$ & $v_\mathrm{ZAMS}\,$ \\
& $M_\odot$ & km/s\\
\hline
MW & 5--35 & 0--350 \\
LMC & 5--45 & 0--350 \\
SMC & 5--50 & 0--350 \\
\hline
\end{tabular}
\end{table}

As outlined in Sect.~\ref{sec2_Cmslairv}, in applying our method we assume that the stellar surface rotation velocity 
remains roughly constant during the main sequence evolution. This is the prediction of the underlying stellar evolution 
models \citep{Brott_2010a} for stars with initial masses below 30\,$M_\odot$. For more massive stars, this feature
still holds as long as the stellar effective temperature is larger than $25\,000\,$K.
For cooler temperatures, mass loss leads to a noticeable spin down of the stars. Our method is therefore valid 
in the ranges given in Table~\ref{tab_valcond}, with a restriction that for stars more massive 
than 30\,$M_\odot$, their effective temperature should be higher than 25\,000\,K.

\section{Application}
\label{sec3}
\subsection{Stellar sample}
The obtained equations to calculate the surface nitrogen abundance for given initial mass and surface rotational velocity 
are applied here to constrain the properties of observed stars under the hypothesis that the underlying stellar models 
describe their evolution correctly. The VLT-FLAMES Survey of Massive Stars \citep{Evans_2005,Evans_2006} has been chosen 
to provide the observational data.

The VLT-FLAMES Survey is a spectroscopic
survey of hot massive stars performed at the {\em Very Large Telescope} of the European Southern Observatory. Over 700 O 
and early B type stars in the Galaxy and the Magellanic Clouds were observed. In \citet{Hunter_2008b} (Magellanic Clouds) 
and \citet{Dufton_2006} (Galaxy) the spectral data has been 
evaluated using the non-LTE TLUSTY model atmosphere code \citep{Hubeny_1995} to gain atmospheric parameters and projected 
rotational velocities of the observed stars.
The surface nitrogen abundances were derived using the nitrogen absorption lines \citep{Hunter_2009}. 
The evolutionary masses of the stars were determined in \citet{Hunter_2008b} using the stellar evolution models of 
\citet{Meynet_1994} and \citet{Schaerer_1993} for LMC objects and in \citet{Meynet_1994} and \citet{Charbonnel_1993} for 
SMC objects. The derived evolutionary masses are in good agreement with evolutionary masses determined from the stellar 
evolution models used here.

For our analysis, we selected stars from the FLAMES LMC and SMC samples by several conditions. It is necessary to know the 
surface nitrogen abundance, the evolutionary mass and the (projected) surface rotational velocity of a star to use our method. 
We disregarded all sample stars for which this is not the case, or for which one of the parameters falls outside the 
validity range of our method (Sect.~\ref{valid}). We further discarded all stars with a projected surface rotational 
velocity below $100\,$km/s, as most of them are expected to be intrinsically slow rotators for which no significant age constraints 
from our method could be expected. In this way, 44 LMC and 35 SMC stars remain on the list, which we call ``our'' sample in the 
following (Tables~\ref{table_dataLMC} and \ref{table_dataSMC}).

While we need to exclude stars with rotational velocities above 350\,km/s, it is possible that stars which rotate faster 
than this are present in our sample, since we only know the projected rotational velocities of our 
sample stars. The largest 
$v \sin (i)$ of all stars in our sample is 323\,km/s. A priori, given the distribution of rotational velocities presented in 
\citep{Brott_2010b} (see Table~\ref{tab_popsynpara}) of stars in the FLAMES LMC and SMC samples as well as a uniform distribution 
of orientations, 4.8\% of all stars with $v \sin (i)<350$\,km/s have surface rotational velocities $v_{\mathrm{rot}}$ higher 
than 350\,km/s. Such fast rotating stars rapidly enrich nitrogen at the surface. Within the first 10\% of their main sequence 
lifetime, their surface nitrogen abundance increases above $\mathcal N = 8\,$dex, in the LMC. For a rough estimate, assume that 
approximately 4.8\% of all stars with $v \sin (i)<350$\,km/s could have a surface rotational velocity above 350\,km/s with 10\% of their 
main sequence lifetime corresponding to a surface nitrogen abundances $\mathcal N \le 8\,$dex. As we have no stars with 
$\mathcal N > 8\,$dex in our sample, 0.48\% of all stars within our sample could have a surface rotational velocity above 350\,km/s, 
which we consider negligible.

The largest $v \sin (i)$ of our stars corresponds to about 50\% of break-up velocity. Consequently, the largest centrifugal 
acceleration in the line of sight amounts to one quarter of the gravitational acceleration. In other words, the measured 
$\log g$ will be decreased by at most about 0.12~dex, and on average by less than half that value. This is smaller than the 
intrinsic measurement error of $\log g$ of about $\pm 0.15$~dex. It is also small compared to the range of gravities of the 
sample stars, which is $4.4\,\mathrm{dex} \leq \log g \leq 3.2$~dex. For this reason, we can neglect the correction of the gravity 
due to rotation.

\subsection{Defining Class 1 stars}
To constrain the age of an observed star through our method, its evolutionary mass, projected surface rotational velocity, 
metallicity and the surface nitrogen abundance need to be known. We want to mention, that in the case of differences between 
evolutionary and spectroscopic masses \citep{Mokiem_2006}, this method will lead to different results 
when using the spectroscopic mass for deriving the nitrogen profile. The effect of changes in mass during the main sequence 
evolution, for stars within the validity range given in Table~\ref{tab_valcond}, on the calculated surface nitrogen abundance 
profile is smaller than the typical observational error of $\pm 0.2$\,dex. The mass, derived from the observational 
data is therefore used as the input value for $M$ in Eq.~(\ref{eq_SNA_1}).

If the true rotational velocity of the star is known (which is rarely the case, and not so for any of our sample stars), 
it can be used as input into Eq.~(\ref{eq_SNA_1}), assuming that it still represents its initial rotational velocity. 
For a given metallicity the surface nitrogen abundance as a function of time is then calculated using the
equations in Appendix~\ref{app1}. Next, the fractional main sequence lifetime is determined by the condition that the 
calculated surface nitrogen abundance matches the observed one. By considering the main sequence lifetime of stars as 
function of mass and surface rotational velocity (cf. Appendix~\ref{sec2_Cmslairv}), the age of the star is obtained.

If the true rotational velocity of a star is not known, but only its projected value $v \sin (i)$, we can compute the 
nitrogen surface abundance as function of $\tau$ for a model with the appropriate mass and metallicity, and with a 
rotational velocity equal to the projected rotational velocity of our target star. Since the projected rotational 
velocity is smaller than the true velocity, the nitrogen enrichment will be slower than in a model rotating with 
the true rotational velocity. We therefore obtain lower limits to the expected nitrogen abundance, for all times. 
The result of the analysis depends now on the measured nitrogen abundance of our star. If it is lower than the largest
computed lower limit (which is obtained for $\tau =1$), we can constrain the upper limit to the current age of the star. 

We define stars of {\em Class~1} by the condition that their {\em observed surface nitrogen abundance, or its
observationally determined upper limit, is lower than the 
maximum (final) calculated surface nitrogen abundance according to Eq.~(\ref{eq_SNA_1}), when adopting the observed 
$v \sin (i)$ as rotational velocity in Eq.~(\ref{eq_SNA_1})}. 

\begin{figure}[htb]
    \centering
        \includegraphics[width=8.5cm]{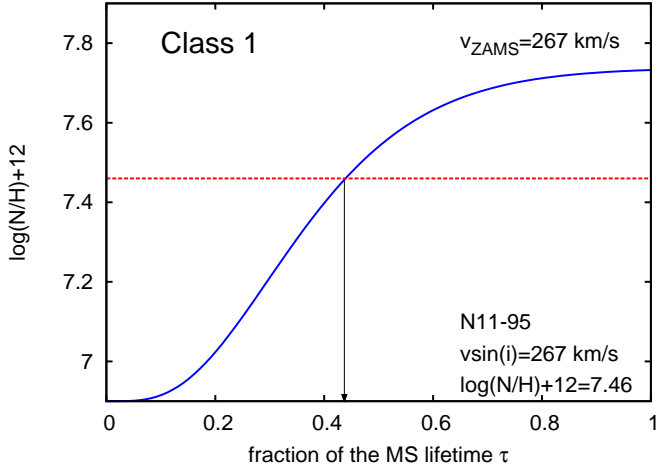}
    \caption{The analysis of the Class~1 stars \object{N 11-95} (LMC),
which has an observed $v \sin (i)$ of 267\,km/s. The blue solid line shows the calculated
surface nitrogen abundance as function of the fractional main sequence life time $\tau$ 
of an LMC star with the mass of \object{N 11-95} and assuming a rotational velocity
of 267\,km/s. The red dotted line indicates the measured surface 
nitrogen abundance of \object{N 11-95}. The black vertical arrow gives the derived upper limit on $\tau$,
i.e. $\tau < 0.43$.}
    \label{fig_class1}
\end{figure}

Figure~\ref{fig_class1} illustrates our method at the example of the star \object{N 11-95} from our LMC sample, with
$M = 15\,M_\odot$, $v\sin(i)=267\,$km/s and $\mathcal N = 7.46\,$dex. The observed abundance of nitrogen at the surface 
equals the calculated nitrogen abundance (which is a function of $\tau$) at a value of $\tau = 0.43$. 
Considering a main sequence lifetime of 11.4~Myr, this leads to an upper age limit of 4.9~Myr for this star.

For some of the sample stars we only know upper limits to the surface nitrogen abundance. Since the nitrogen upper limit is 
larger than the true nitrogen abundance of the sample star, using the upper limit to the nitrogen abundance leads to an 
upper age limit of the observed star in the same way as for stars with determined nitrogen abundances.

\subsection{Population synthesis of Class~1 stars}
It is interesting to consider the location of Class~1 stars in the so called Hunter-diagram,
where the surface nitrogen abundance of stars is plotted versus their projected rotational velocity.
Since according to the stellar evolution models fast rotators are achieving higher surface nitrogen 
abundances, stars are likely to fall into Class~1 when they have a large $v \sin(i)$ (which makes
their maximum nitrogen abundance large) as long as their measured nitrogen abundance is not too high.
Conversely, nitrogen rich apparent slow rotators will not fall into Class~1. 

To study this, we perform a population synthesis simulation for rotating single massive main sequence stars
of LMC metallicity using the code {\sc STARMAKER} \citep{Brott_2010b}, with parameters as summarized in 
Table~\ref{tab_popsynpara}.

\begin{table}[h]
\caption{Parameters used in our population synthesis calculation (cf. Fig.~\ref{fig_popsyn}).}
\label{tab_popsynpara}
\centering
\begin{tabular}{l l}
\hline\hline
parameter &  \\
\hline
star formation rate & constant\\
considered time interval & 35~Myr\\
velocity distribution & Gaussian distribution \citep{Brott_2010b} \\
& ($\sigma=141\,$km/s, $\mu$=100\,km/s)\\
velocity range & 0 -- 570\,km/s\\
mass distribution & \citet{Salpeter_1955} IMF\\
mass range & 5 -- 30\,$M_{\odot}$\\
stellar model grid & LMC \citep{Brott_2010a} \\
\hline
\end{tabular}
\end{table}

To verify the Class~1 character of a simulated star with a given age, its mass and
$v\sin(i)$ value have been used to compute its surface nitrogen abundance at the end of its main sequence evolution according
to Eq.~(\ref{eq_SNA_1}). If this value is higher than its actual surface nitrogen abundance, it is
taken to be of Class~1. As all stars in the LMC and SMC sample have masses below 30\,$M_{\odot}$ the
maximum mass considered in this simulation was chosen as 30\,$M_{\odot}$. Only stars with $v\sin(i) > 100\,$km/s are evaluated.

\begin{figure}[htb]
    \centering
        \includegraphics[width=8.5cm]{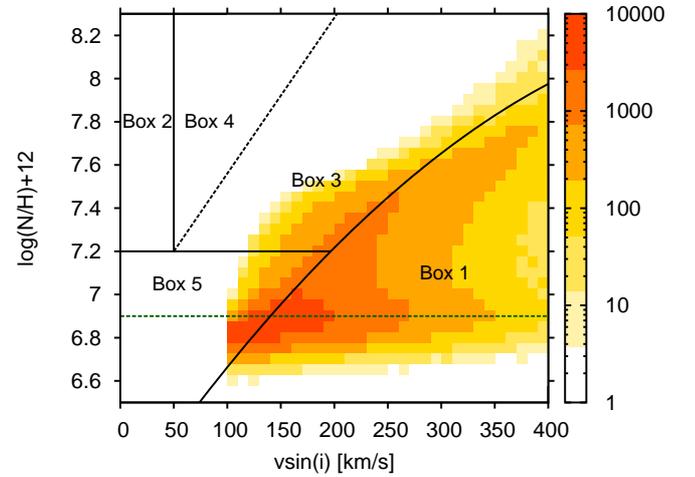}
    \caption{Hunter-diagram of Class~1 stars of a synthetic population of main sequence stars obtained using the {\sc STARMAKER} 
code \citep{Brott_2010b}. The surface nitrogen abundance is plotted versus the projected surface rotational velocity. The colors 
indicate the number of Class~1 stars in a given pixel. The absolute scaling is arbitrary. The Hunter-diagram is divided into 5 
areas as in \citep{Brott_2010b} to simplify the later analysis.}
    \label{fig_popsyn}
\end{figure}

Figure~\ref{fig_popsyn} shows the number density of simulated Class~1 stars in the Hunter-diagram. 
Stars are counted within intervals of
$\Delta v \sin(i) = 10\,$km/s and $\Delta \mathcal N = 0.05\,$dex. The number of models in the intervals are indicated by the
color coding.

Figure~\ref{fig_popsyn} illustrates for which stars the nitrogen chronology method can be used to derive an upper age limit.
As expected, preferentially stars in the lower right corner
of the Hunter-diagram can be evaluated as Class~1. For stars in the upper left corner of the Hunter-diagram we will 
be able to constrain their inclination angles (see below). Finally, we note that if the inclination angle of 
a star is observationally determined (e.g. through a pulsation analysis, or a binary membership), its nitrogen age
can be determined regardless of its location in the Hunter-diagram.

The majority of Class~1 stars in our sample are located in the LMC, only one star belongs to the SMC sample 
(see Appendix~\ref{app2}). The reason is that nitrogen abundances of SMC stars are lower than those of comparable LMC
(or Galactic) stars \citep{Hunter_2009}, such that their abundance can not be determined if it is below a certain
threshold which increases with the projected rotation velocity. That means, Box~1 in the Hunter-diagram of the
SMC early B~star sample is essentially empty (See Fig.~7 in \citet{Hunter_2009}). 

Furthermore, a comparison of Fig.~\ref{fig_popsyn} with Fig.~9 of \citet{Brott_2010b} shows that only very few stars 
from the considered sample are expected in Box~1, due to the visual magnitude cut-off imposed in the sample selection. 
For this reason, \citet{Brott_2010b} concluded that those stars of this sample which do appear in Box~1 are not evolving 
as rotationally mixed single stars.

\subsection{Defining Class~2 stars}
\label{Sec_def_C2}
We define {\em Class~2} stars by the condition that the {\em observed nitrogen abundance,
or its observationally derived upper limit, is higher than the maximum (final) 
calculated surface nitrogen abundance according to Eq.~(\ref{eq_SNA_1}) when adopting 
the observed $v \sin(i)$ as rotational velocity in Eq.~(\ref{eq_SNA_1})}. 
Stars are thus either of Class~1 or of Class~2.
For Class~2 stars, the derived upper limit to the current age is the main sequence lifetime, which yields no age constraint.

However, for a Class~2 stars with a determined nitrogen abundance, 
we can find the minimum required rotational velocity $v_\mathrm{min}$ 
to achieve this nitrogen value for the given mass of the star. Since the surface nitrogen abundance is monotonically 
increasing with time, the observed nitrogen abundance will then be produced by a model star at the end of the main 
sequence evolution ($\tau = 1$), rotating with this minimum velocity.
By using the minimum rotational velocity, the maximum inclination angle which is compatible with the observed nitrogen 
abundance, which we designate as nitrogen inclination $ i_\mathrm{N} $, can be derived for the star. 
The sine of $ i_\mathrm{N} $ can be calculated as 
\begin{equation}
\sin \left( i_\mathrm{N} \right) = \frac{v \sin \left( i \right)}{v_\mathrm{min}}.
\label{eq_max_incl}
\end{equation}

\begin{figure}[htb]
    \centering
        \includegraphics[width=8.5cm]{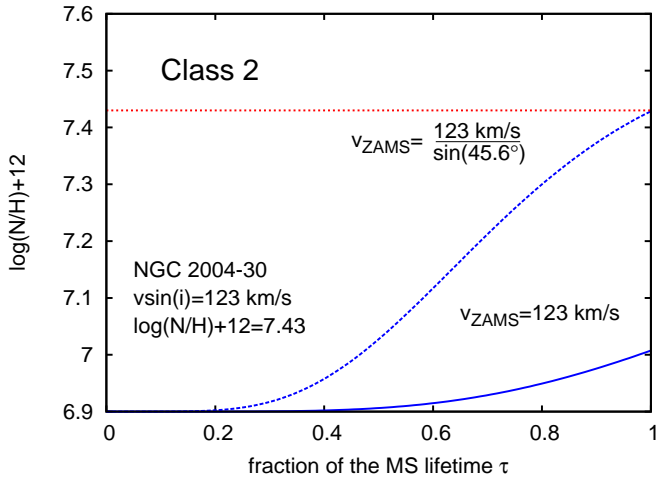}
    \caption{The determination of the nitrogen inclination angle of a Class~2 star is shown for the example \object{NGC 2004-30}. The 
red dotted line indicates the observed surface nitrogen abundance and the blue solid line represents the calculated 
surface nitrogen abundance as a function of the fraction of the main sequence lifetime using $v_\mathrm{ZAMS} = 123\,$km/s 
in Eq.~(\ref{eq_SNA_1}). The blue dashed line is calculated with a value of the surface rotation velocity of 
$v_\mathrm{ZAMS} = 173\,$km/s. This corresponds to an inclination angle of $\sin (i_\mathrm{N})= 0.71$.}
    \label{fig_class2}
\end{figure}

Figure~\ref{fig_class2} illustrates the analysis of Class~2 stars at the example of the LMC star \object{NGC 2004-30}. 
To calculate the evolution of the surface nitrogen abundance, $M = 19\,M_\odot$ and $v_\mathrm{ZAMS} = 123\,$km/s 
have been used. The observed nitrogen value of $\mathcal N = 7.43\,$dex 
is higher then the predicted one during the whole main sequence evolution. 
The observed nitrogen value can be reached with a minimum rotational velocity of $v_\mathrm{min} = 172\,$km/s, which 
leads to $\sin (i_\mathrm{N})=0.71$ (blue dotted line), i.e. $i_\mathrm{N}=45.6^{\circ}$. 

For Class~2 stars where only an upper limit to the surface nitrogen abundance has been determined,
a nitrogen inclination can be determined in the same way, by adopting the upper limit as nitrogen abundance.
However, in this case, the nitrogen inclination does not provide an upper limit to the true
stellar inclination, since lower nitrogen surface abundances would lead to larger nitrogen inclinations.
Their values are nevertheless interesting, as shown in Sect.~5 for the Class~2 stars in the SMC.

\section{Results: Class 1 stars}
\label{sec4}
\subsection{Isochrone ages}
To evaluate the age constraints obtained for Class~1 stars via the nitrogen abundances they are compared to the results 
from the well proven method of estimating the age of a star through its position in the Hertzsprung-Russel (HR) diagram. The 
idea is to compare effective temperature and luminosity of the star with isochrones plotted in the HR diagram.

\begin{figure}[htb]
    \centering
        \includegraphics[width=8.5cm]{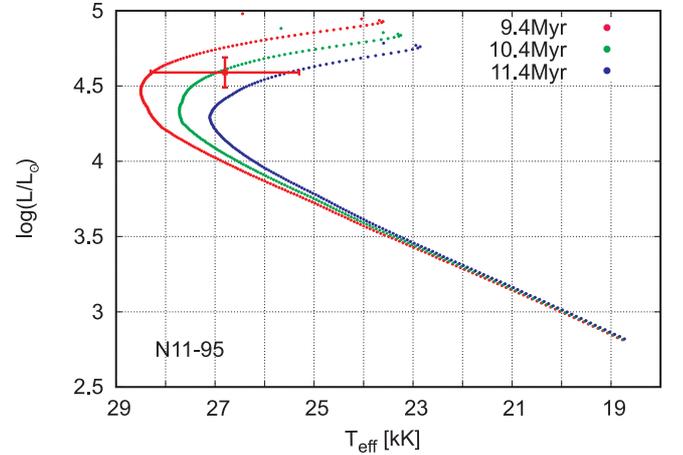}
    \caption{Isochrone fitting in the HR diagram for the LMC star \object{N 11-95}, which is indicated by the red point with error bars. 
The red, green and blue lines are isochrones from non rotating stellar models for different ages. 
The green line defines the most likely age of the star while red and blue represent maximum and minimum possible 
current age when considering uncertainties in measured effective temperature and luminosity of the star.}
    \label{fig_isochones}
\end{figure}

We computed isochrones using non-rotating models of the stellar evolution grids of \citep{Brott_2010a} up to an age of 50~Myr 
with a step size of 0.2~Myr. Comparing the isochrones with the position of a star the best fitting isochrone 
defines its isochrone age $t_\mathrm{isochrone}$ as shown exemplary for the star \object{N 11-95} in Fig.~\ref{fig_isochones}. 
The observational error bars imply an uncertainty in the isochrone ages of about $\pm$ 1~Myr.

All stars of Class~1 have been analyzed this way to obtain their isochrone age.  
Since non-rotating models were used, rotational mixing is not considered in estimating the isochrone ages of the stars.
However it leads only to negligible errors in the mass and rotational velocity range considered here (see 
Fig.~\ref{fig_frac_MS_lifetime} in Appendix~\ref{app1}).
All isochrone ages are given in Tables~\ref{table_dataLMC} and \ref{table_dataSMC}.

Assuming that the isochrone age reproduces the current age of a star well, our method allows to derive its 
inclination angle. Due to uncertainties in the isochrone age $t_\mathrm{isochrone}$, and when using Eq.~(\ref{eq_tauMS}) to 
derive the main sequence lifetime $\tau_\mathrm{H}$, 
it is possible that $\tau_\mathrm{H} < t_\mathrm{isochrone}$. In this case we limit 
$t_\mathrm{isochrone} = \tau_\mathrm{H}$ (occurred in 3 out of 16 cases). 
To derive the inclination angle, the isochrone age needs to be smaller 
than the upper limit to the current age using the nitrogen abundance.
Amongst the stars of Class~1 (the method is similar to deriving the inclination angle for Class~2 stars, see 
Sect.~\ref{Sec_def_C2}), a few stars fulfill this condition. The derived inclination angles are given in 
Tables~\ref{table_dataLMC} and ~\ref{table_dataSMC}. 
In particular, we find that the most likely inclination angles of stars L15 (\object{N 11-120}) and L42
(\object{NGC 2004-107}) are 66$^\circ$ and 62$^\circ$, respectively, while L6 (\object{N 11-89}) and L8
(\object{N 11-102}) are seen nearly equator-on.

\subsection{Nitrogen ages versus isochrone ages}
\begin{figure}[htb]
    \centering
        \includegraphics[width=8.5cm]{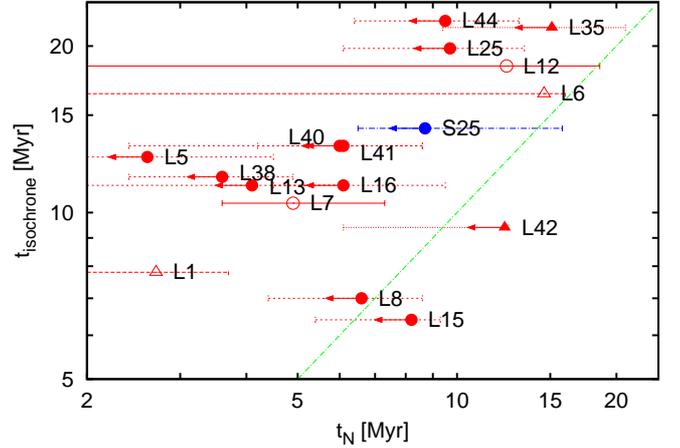}
    \caption{The upper age limits $t_N$ obtained from nitrogen chronology for Class~1 stars are plotted 
against the ages derived  using isochrone fitting in the HR diagram, $t_{\mathrm{isochrone}}$. Stars located in the LMC are 
shown with red symbols while the SMC star is depicted in blue. Stars for which only upper limits to the surface nitrogen 
abundances could be determined are marked by filled 
symbols with arrows indicating the direction of change in the results when the surface nitrogen abundance is smaller 
then the value of the upper limit. Circles mark single stars and triangles indicate
probable binaries. The green dotted line is defined by $t_{\mathrm{N}} = t_{\mathrm{isochrone}}$.} 
    \label{fig_comparison}
\end{figure}

The results of the age constraints through nitrogen are summarized in Tables~\ref{table_dataLMC} and
\ref{table_dataSMC} for all considered stars. 
Fig.~\ref{fig_comparison} shows the upper age limits of our sample stars
obtained via nitrogen chronology $t_{\rm N}$, plotted against the ages obtained through isochrone fitting, 
$t_{\mathrm{isochrone}}$. Only stars of Class~1 are depicted.
Due to the distinction between Class~1 and Class~2, the stars in Fig.~\ref{fig_comparison} are those which have 
preferentially low surface nitrogen abundances and high projected surface rotational velocities. 
Figure~\ref{fig_HRD} shows the distribution of Class~1 stars in the HR~diagram.

Since the projected rotational velocity provides a lower limit to the true rotational
velocity of a star, the surface nitrogen abundance results in an upper age limit, i.e., 
the star is younger or as old as the obtained nitrogen age. Therefore, the upper age limit derived from the surface nitrogen 
abundance should be larger than the age determined using isochrones,  under the assumption that the age of 
a star is approximated well by the second method. This means that stars should be placed below the green dotted line
in Fig.~\ref{fig_comparison}, which corresponds to $t_{\mathrm{isochrone}} = t_{\mathrm{N}}$. 

Considering that the error in mass is of the order of 10-25\% \citep{Mokiem_2006} --- which is insignificant
given the weak mass dependence of the nitrogen enrichment in the considered mass range (Fig.~\ref{fig_nitrogen}) ---
and the error in $v \sin (i)$ is only about 10\% \citep{Hunter_2008b}, the error of $\pm 0.2$ dex for the surface nitrogen
abundance \citep{Hunter_2009} is indeed the dominant one.
To estimate its influence, we repeat the analysis for
each star, comparing the surface nitrogen abundance according to Eq.~(\ref{eq_SNA_1}) with an abundance which is
0.2\,dex higher or lower than the measured surface nitrogen abundance. The lower nitrogen abundance leads to a
decrease while the upper value results in an increase of the upper age limit for the star.
The uncertainties of the upper age limit due to the uncertainties in the surface nitrogen abundance are given in
Tables~\ref{table_dataLMC} and \ref{table_dataSMC}, and are plotted as horizontal error bars in Fig.~\ref{fig_comparison}.

The majority of our sample stars are placed above the green dotted line in Fig.~\ref{fig_comparison}, even
when considering the error in their measured surface nitrogen abundance. 
This contradicts the expected behaviour. 
For example, the star \object{N 11-95} (star L7 in our sample, see Table \ref{table_dataLMC}) analyzed in Fig.~\ref{fig_class1} has
an upper age limit from nitrogen
chronology of 5.8~Myr. However, its isochrone age is 10.4~Myr. This means that, according to its position in the HR
diagram, this rapidly rotating star ($v \sin (i) = 267\,$km/s) is seen near its TAMS position in the HR diagram,
but it has only a modest surface nitrogen enhancement.

Only two out of 17 stars are found in the predicted area (\object{N 11-120}$=$L15, and 
\object{NGC 2004-107}$=$L42).
Considering the uncertainty of $\pm$0.2\,dex in the measured surface nitrogen abundance, we see in Fig.~\ref{fig_comparison}
three more stars (\object{N 11-102}$=$L8, \object{N 11-116}$=$L12, and \object{NGC 330-51}$=$S25) with an error bar reaching 
into the allowed area, and two further stars (\object{N 11-89}$=$L6 and \object{NGC 2004-88}=L35) with an
error bar getting very close to the allowed area. 

\begin{figure}[htb]
    \centering
        \includegraphics[width=8.5cm]{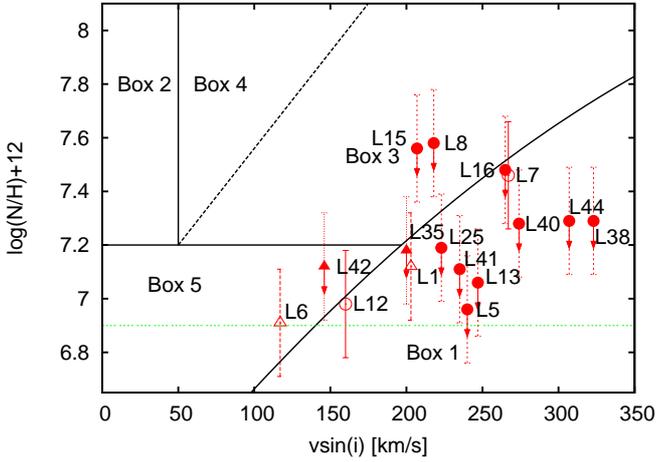}
    \caption{Hunter-diagram of all stars shown in Fig.~\ref{fig_comparison}. The surface 
nitrogen abundance is plotted as a function of the projected surface rotational velocity for stars located in 
the LMC. Upper limits to the surface nitrogen abundance are marked by 
filled symbols. In addition there is a differentiation between single stars (circles) and probable binaries 
(triangles). A subdivision into five boxes as defined in \citet{Brott_2010b} has been done to simplify the analysis. The 
green dotted line represents the initial surface nitrogen abundance for LMC B stars.}
    \label{fig_hunter}
\end{figure}

\begin{figure}[htb]
    \centering
        \includegraphics[width=8.5cm]{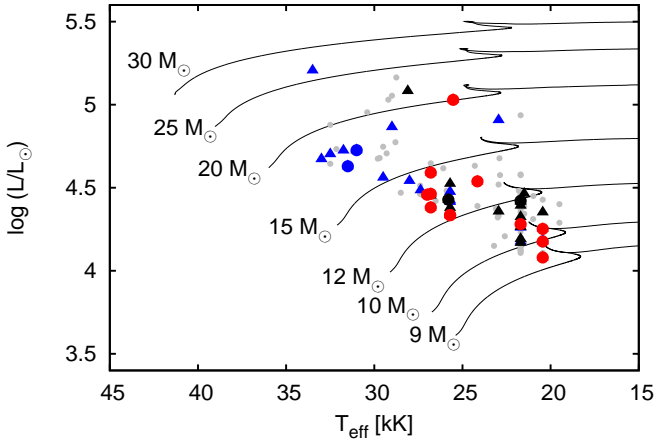}
    \caption{The Hertzsprung-Russel diagram is shown for several evolutionary models (LMC) with initial masses from 9 
to 30\,$M_{\odot}$ and surface rotational velocities of approximately 200\,km/s. Additionally, our sample with LMC-metallicity 
is depicted. Stars of Class~1 are indicated by circles and stars of Class~2 are represented by triangles. With respect to 
Fig.~\ref{fig_hunter}, stars in Box~1 are shown in red, star in Box~3 in blue and stars in Box~5 in black. Grey dots depict 
the remaining not analyzed stars of the VLT-FLAMES Survey (LMC) after the selection criteria applied in 
\citet{Brott_2010b}.}
    \label{fig_HRD}
\end{figure}

\subsection{Discussion}
As shown above, the nitrogen chronology clearly fails for 10 out of the 17 Class~1 stars in our sample.  
It is interesting to consider the location of these stars in the Hunter-diagram (Fig.~\ref{fig_hunter}).
It turns out that all 10 stars are located in Box~1. In contrast, amongst the 7~stars for which 
the nitrogen chronology might work, the six LMC stars are found in Box~3 (L8, L15), in Box~5 (L6, L42),
or in Box~1 but very near Box~5 (L12, L35). The SMC star S25 can not be compared in the same Hunter-diagram. 
For the LMC Class~1 stars, also the reverse statement holds: Nitrogen chronology might work for all stars
which are not in Box~1, or which are in Box~1 very near Box~5. In other words, a slight shift of the 
dividing line between Box~5 and Box~1 would result in separating the Class~1 stars for which
nitrogen chronology clearly fails (Box~1) from those for which it might work (outside of Box~1).

The isochrone age of the 10 LMC stars in Box~1 for which nitrogen chronology fails
is larger than the upper age limit derived from their surface nitrogen abundance, 
given their rather high projected rotational velocities. Due to the simplicity of our 
method the revealed conflict can only be due to the stars in Box~1 behaving differently than predicted by the 
evolutionary models for rotationally mixed single stars of \citet{Brott_2010a}. 

The detailed and tailored population synthesis of the LMC early B~type main sequence stars from the FLAMES Survey by
\citet{Brott_2010b} already pointed out that the stars in Box~1 could, in a statistical sense, not be
reproduced through the stellar models of single stars including rotational mixing.
Our nitrogen chronology analysis is not a statistical method, but rather considers each star individually.
While it uses the same underlying stellar evolution models as \citet{Brott_2010b}, the latter did not consider age
constraints. In this respect, our chronology method provides a new test of the 
evolutionary models provided by \citet{Brott_2010a} which is independent of the results of \citet{Brott_2010b}. 
Nevertheless, our result is very much in line with that of
\citet{Brott_2010b}, i.e., that Box~1 stars in our sample are generally not consistent
with the models of \citet{Brott_2010a}. 

While the errors due to inaccuracies of our fitting method are mostly insignificant compared to
the observational errors in nitrogen, the main theoretical error source could be
the systematic error introduced by the physics included in the underlying stellar evolution models.
However, the mixing physics (convective overshooting and rotational mixing) in these models
was specifically calibrated using the LMC sample of the FLAMES Survey.
 
We are left with the conclusion that either rotational mixing does not operate in massive stars,
or that the stars in Box~1 of the Hunter-diagram (Fig.~\ref{fig_hunter}) are affected by physics
which is not included in the stellar evolution models (e.g., binary effects).
In this context, we point out that while rotating stars may well evolve into Box~1
in general (cf. Fig.~\ref{fig_popsyn}), most of them would not be detected in the FLAMES survey, due to the employed
magnitude cut-off (cf. Figs.~6 and~10 of \citet{Brott_2010b}). The origin of these stars is therefore in question.

For example, it has been shown that binary evolution can place stars into Box~1 in the Hunter-diagram
\citep{Langer_2008,Mink_2011} for which the age constraints derived from nitrogen chronology do not apply.
However, while four Class~1 stars in our sample are likely binaries (Fig.~\ref{fig_comparison}),
the remaining twelve show no indication of binarity, despite the fact that multi epoch spectra are 
available for them. The FLAMES Survey observing campaign was not tailored for binary detection. Applying
the method developed by \citet{Sana_2009} the limited number of epochs should roughly result in a detection
probability of at least 80\% for systems with periods 
up to 100 days, dropping rapidly to a detection rate of less than 20\% for periods
longer than a year. Furthermore, while some 
binaries amongst the Box 1 stars may remain undetected, it is the pre-interaction
binaries which are more likely to be detected \citep{Mink_2011}. 
With the data at hand, it therefore seems unlikely that we can evaluate the
possibility that binarity by itself can completely explain 
the situation.
 
When we apply nitrogen chronology to LMC stars outside of Box~1 in Fig.~\ref{fig_hunter}
--- i.e., to the only stars which can be expected to be compatible with the single star models
of \citet{Brott_2010a} --- the result is positive. While there are only four stars (L6, L8, L15, L42),  
their nitrogen age is consistent with their isochrone age:
They are either in the allowed part of Fig.~\ref{fig_comparison} (L42, L15), or nearest to 
its borderline (L6, L8). As explained above, a slight shift in the borderline between Boxes~1 and~5
in Fig.~\ref{fig_hunter} includes two more stars (L12, L35) into this consideration.
Together with the SMC star S25, for which can not be considered in Fig.~\ref{fig_hunter}, there
are seven stars for which the nitrogen chronology yields promising and consistent results.

In summary, we conclude that our new method give results for Class~1 stars which
are consistent with those from classical isochrones for all stars in the subgroup where this could be expected.
Furthermore, it flags contradictions for all stars in the second subgroup which, according to \citet{Brott_2010b}, 
are not thought to have evolved as rotationally mixed single stars. 
For stars of both subgroups, the application of nitrogen chronology is fruitful
and provides insights which are not obtainable otherwise.

Obviously, it would be desirable to apply nitrogen chronology to
a stellar sample where the observational
bias does not exclude Box~1 of the Hunter-diagram to be populated by rotationally 
mixed single stars --- which according to Fig.~\ref{fig_popsyn}
appears indeed possible. While a
population of stars which contradict the underlying stellar models (presumably binaries)
might still be present in such a sample, 
expectedly it would be diluted in a dominant population of single stars.

\section{Results: Class 2 stars}
\label{sec5} 
Stars are of Class~2 if the observed surface nitrogen
abundance is higher than the predicted value of a star rotating with an unprojected velocity
equal to the $v \sin(i)$ of our star at the end of the main sequence.
This is the case for most of our sample stars residing in Box~3 of the Hunter-diagram (see Fig.~\ref{fig_hunter}),
which according to the statistical analysis of \citet{Brott_2010b} agree reasonably
well with the single star evolution models including rotational mixing.
None of the LMC Class~2 stars is located in Box~1 of the Hunter-diagram. 
While we can not derive any age constraints,
we can here, relying on single star models with rotational mixing, constrain their
inclination angles.

\begin{figure*}[htbp]
    \centering
        \includegraphics[width=17cm]{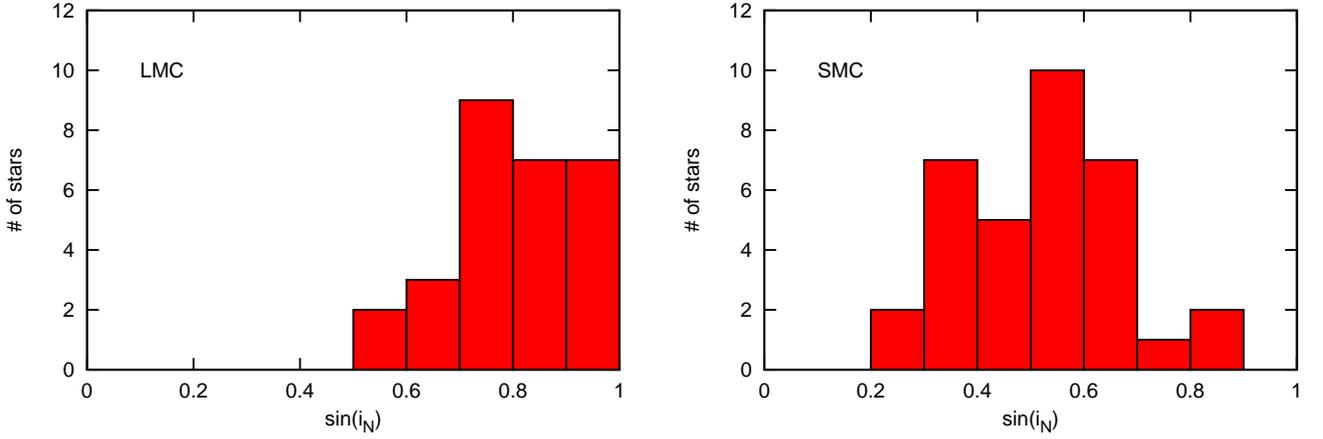}
    \caption{Histograms of the $\sin (i_{N})$-values for the Class~2 stars of our sample located in the LMC (left panel) 
and SMC (right panel). The number of stars within an area of $\Delta \sin (i_{N})= 0.1$ were counted.}
    \label{fig_class2_2}
\end{figure*}

The results of the analysis for our sample are summarized in Tables~\ref{table_dataLMC} and \ref{table_dataSMC}, 
where we give the
upper limits on the inclination angle for all stars. Figure~\ref{fig_class2_2} depicts the histograms for the 
nitrogen inclination angles for the 28~LMC and 24~SMC Class~2 stars of our sample. Assuming random orientations of stars, 
the distribution of $\sin (i)$ increases for increasing values of $\sin (i)$. 
Since we derive upper limits on the inclinations for stars with determined surface nitrogen abundance, 
the distribution of those can be expected to be shifted to even higher values.
 
The expected behaviour is seen for our LMC stars, where the distribution of the sine of the nitrogen inclination
angles favours high values with a maximum between 0.7 and 0.8. We refrain from a further analysis of this histogram since the
selection of Class~2 stars introduces a bias against the fastest rotators (see Fig.~\ref{fig_popsyn}).
While the number of very fast rotators is small, this introduces a bias in our distribution
which is difficult to account for.
 
However, it is interesting to note that the
distribution of the nitrogen inclination angles of the
SMC stars shows a different behaviour. Here, high values are rare and
the average $\sin (i_\mathrm{N})$ value is about 0.5. 

For most SMC stars, only an upper limit to the surface nitrogen
abundance is known. The higher his limit, the lower is the
$\sin (i_\mathrm{N})$ value required to reach this value. 
Thus, if the true surface nitrogen abundance is below the value of the upper limit,
the $\sin (i_\mathrm{N})$ value derived from the upper limit will be
smaller than the one which would be derived from the true nitrogen abundance. 
This implies that the $\sin (i_\mathrm{N})$ value derived from an upper limit on nitrogen
may be smaller than the true $\sin i$-value of the star. 
Therefore the maximum, or mean, of the $\sin (i_\mathrm{N})$ distribution
can be located at lower $\sin (i)$-values than the maximum of the true $\sin (i)$ distribution.  

Interpreting the distribution of nitrogen inclinations of the SMC stars in Fig.~\ref{fig_class2_2}
in this way, and assuming randomly oriented rotational axes of these stars, 
we conclude that the nitrogen upper limits derived for many of the
SMC stars are significantly larger than their true nitrogen abundances.
Considering Fig.~7 of \citet{Hunter_2009}, which compares the location of the
SMC stars with evolutionary tracks in the Hunter-diagram, one can see that
such a significant shift would indeed be required to bring the observations into
agreement with the models. 
 
Similar to the analysis of Class~1 stars, we can use the isochrone ages of Class~2 stars with determined nitrogen
abundance to derive their inclination angles.
If only an upper limit to the surface nitrogen abundance is known,  
a lower limit to the true inclination angle can be found. 
These inclination angles are designated as $i_\mathrm{isochrone}$ and their sine is given in 
Tables~\ref{table_dataLMC} and \ref{table_dataSMC} for all sample stars. 
 
\section{Conclusions}
\label{sec6}
We present a new method of nitrogen chronology, which can constrain the age of a star, and/or its inclination angle,
based on its observed surface nitrogen abundance, mass and projected surface rotational velocity,
by comparing the observed nitrogen abundance with the one predicted by the theory of rotational mixing in single stars.

This method can be applied to stars when their surface nitrogen abundance increases
monotonically with time during the main sequence evolution.
It has been worked out here for stars in the mass range between 
5\,$M_{\odot}$ and up to 35--50\,$M_{\odot}$, and for metallicities adequate for the Milky Way, the LMC and the SMC,
based on the stellar evolution models of \citet{Brott_2010a}.
We apply our method to 79 stars from the early B~type LMC and SMC samples of the VLT-FLAMES Survey of Massive Stars
\citep{Evans_2005,Evans_2006}. 

Age constraints from nitrogen chronology could be obtained for 17~of the 79 analyzed stars. 
In Sect.~4.2, we compared those to ages of these stars as derived from isochrone
fitting in the HR diagram (Fig.~\ref{fig_comparison}). We found the isochrone ages of 10 objects to be incompatible
with their nitrogen age constraints. Based on their rapid rotation and low
nitrogen enrichment, \citet{Brott_2010b} concluded that these 10~stars did not
evolve as rotationally mixed single stars. Our star-by-star analysis of these objects is 
reinforcing their conclusion that the weakly enriched fast rotators in the
LMC early B~stars of the FLAMES Survey --- which are 15\% of the survey stars --- can not
be explained by rotating single stars. 

For the remaining 7~stars with nitrogen age constraints, six LMC and one SMC star,
nitrogen and isochrone ages are found to be consistent within error estimates.
Four of them show good agreement, which also allows to derive their inclination angles (Sect.~4.2).

We conclude that our nitrogen chronology results are consistent with the isochrone method
for stars which have no indication of an unusual evolution, but incompatible for stars  
which are suspected binary product according to \citet{Brott_2010b}. Therefore, our new method 
provides important new results for both groups of stars. However, it is clearly desirable to
obtain larger samples of stars especially of the first group. 

For the 62~stars (28 LMC and 34 SMC stars) for which no age constraint could be derived, 
our method provides limits on the inclination for each object (Sect.~5).
While the results for the LMC stars are roughly consistent with random inclination angles,
this appears to be different for the SMC stars, for which only upper limits
to the nitrogen abundance are available in most cases \citep{Hunter_2009}. We argue that random inclinations
are also realized for the SMC stars, with the implication that the true nitrogen surface abundances
of most SMC stars are significantly below the derived upper limits.

We believe that our new method can be used in the near future to
provide further tests of the theory of rotational mixing in stars,
and --- if such tests converge to confirm this theory --- to act as a new chronometer
to constrain the ages of massive main sequence stars.

Based on the method presented here, a web tool\footnote{http://www.astro.uni-bonn.de/stars/resources.html} 
can be found online. For given mass, (projected) surface rotational velocity 
and metallicity, the surface nitrogen abundance is calculated and compared to the observed surface nitrogen abundance to 
constrain the age and inclination angle. 

\begin{acknowledgements} 
We acknowledge fruitful discussions with Hugues Sana, and are grateful to an anonymous referee
for helpful comments on an earlier version of this paper.
\end{acknowledgements}
 
\begin{appendix}
\section{Equations}
\subsection{Determining parameters $a$, $b$, $c$ and $d$ (Eq.~(\ref{eq_SNA_1})) for the example of the LMC metallicity}
\label{app1}
In this section, we derive the parameters $a$, $b$, $c$ and $d$ of Eq.~(\ref{eq_SNA_1}). 
This is done for the example of the LMC grid of stellar evolution models presented in \citet{Brott_2010a}.

Our dataset of the stellar evolution models contains information about the time-evolution of the surface nitrogen abundance. 
We can derive the main sequence lifetime from our models and therefore obtain the surface nitrogen abundance as a function of 
the fraction of the main sequence lifetime. 

The parameter $a$ corresponds to the initial surface nitrogen abundance and is set to be the surface nitrogen 
abundance at $\tau=0$, which for the LMC is given by the constant value
\begin{equation}
a = 6.9. 
\label{eq_paraa_comp}
\end{equation}

By fitting Eq.~(\ref{eq_SNA_1}) to the data of the stellar evolution models of the LMC grid, we obtain the remaining 
three parameters for different masses and surface rotational velocities at the ZAMS.

\begin{figure}[htbp]
    \centering
        \includegraphics[width=8.5cm]{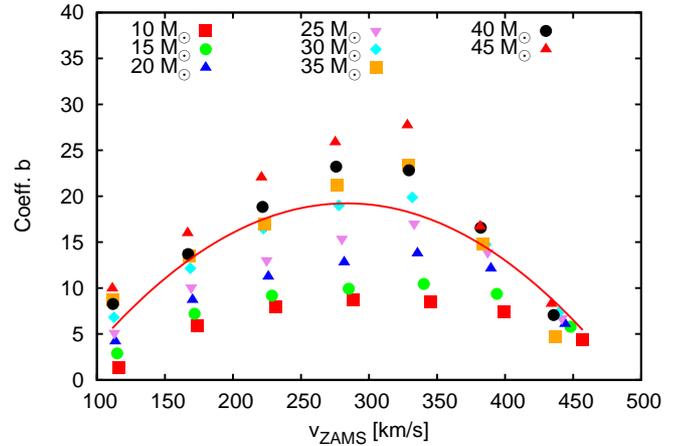}
    \caption{The parameter $b$ is shown as a function of initial surface rotational 
velocity for several different masses, obtained by fitting Eq.~(\ref{eq_SNA_1}) to the 
stellar evolution grid data (LMC). The solid line represents the best fit for the 30\,$M_\odot$ models.}
    \label{fig_coeffb}
\end{figure}

In Fig.~\ref{fig_coeffb}, the fitted parameter $b$ is shown as a function of $v_\mathrm{ZAMS}$ for several different initial 
masses. A polynomial function of second order reproduces the data for fixed mass well. The parameter $b$ is therefore 
calculated as
\begin{eqnarray}
b = b_1 \cdot v_\mathrm{ZAMS}^2 + b_2 \cdot v_\mathrm{ZAMS} + b_3. \nonumber
\end{eqnarray}
\begin{figure*}[htbp]
    \centering
        \includegraphics[width=17cm]{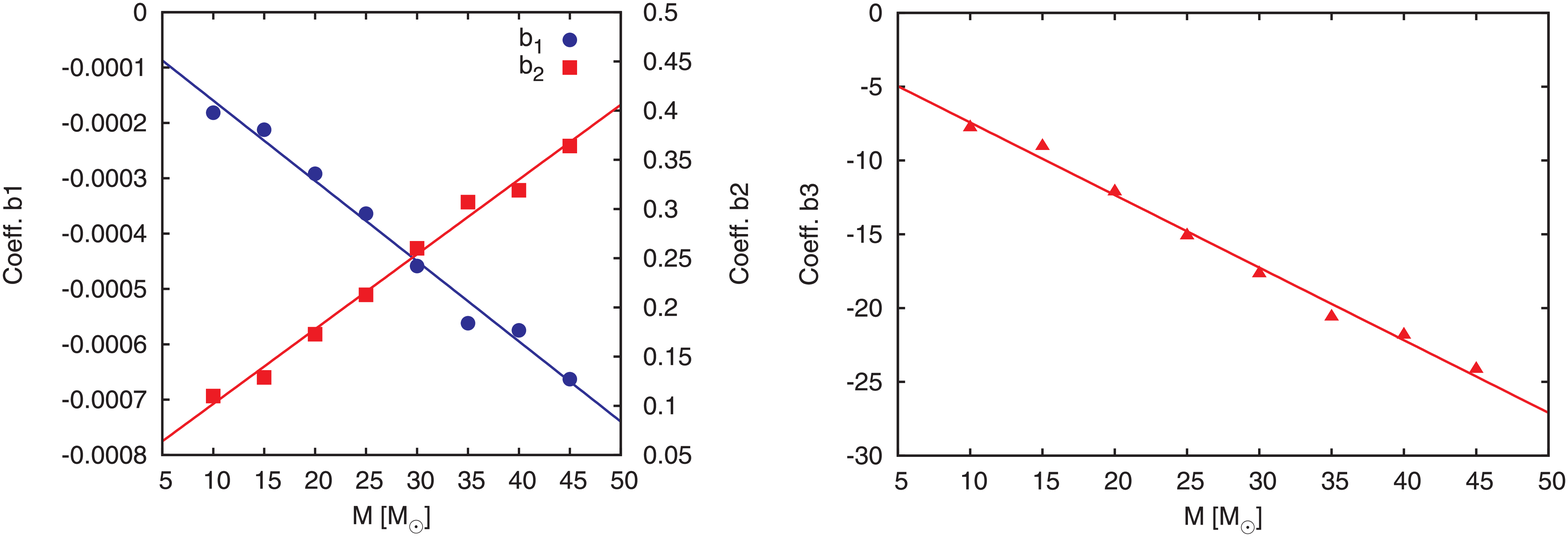}
    \caption{The three coefficients $b_1(M)$, $b_2(M)$ and $b_3(M)$ of the polynomial function to describe the parameter 
$b$ are depicted as functions of mass. All parameters can be fitted well using linear functions which are 
plotted by solid lines.}
    \label{fig_coeff_b}
\end{figure*}

Figure~\ref{fig_coeff_b} shows the coefficients $b_1$, $b_2$ and $b_3$ as a function of initial mass. They can be fitted well by 
linear approximations shown by the solid lines. Using $b_1(M)$, $b_2(M)$ and $b_3(M)$ the 
parameter $b \left( M \left[ M_\odot \right], v_\mathrm{ZAMS} \left[ \mathrm{km/s} \right] \right)$ can be calculated 
using
\begin{equation}
b =  - \frac{M +4.6}{2.0} + \left( M + 3.2 \right) \cdot \frac{v_\mathrm{ZAMS}}{131} - \left( M + 0.8 \right) \cdot \left( \frac{v_\mathrm{ZAMS}}{263} \right)^2.
\label{eq_parab_comp}
\end{equation}

\begin{figure*}[htbp]
    \centering
       \includegraphics[width=17cm]{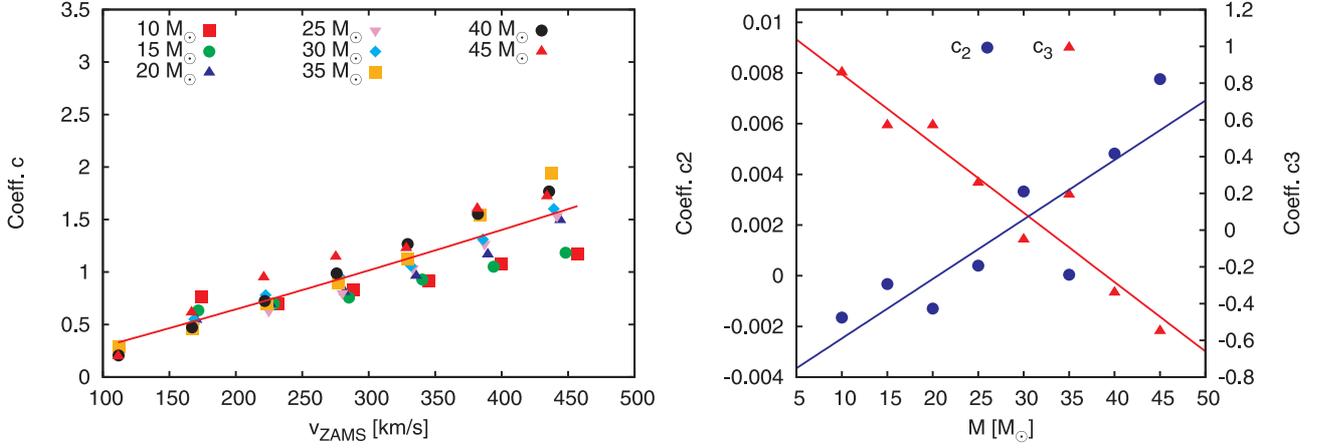}
    \caption{On the left panel, the parameter $c$ is shown as a function of initial surface rotational 
velocity for several different masses. The line represents the best fit for the 30\,$M_\odot$ model. On the right panel, the coefficients  
$c_2(M)$ and $c_3(M)$ are plotted as functions of the initial mass. They can be reproduced best using linear 
functions shown by solid lines.}
    \label{fig_coeff_c}
\end{figure*}

Similarly we obtained parameter $c$ as a function of initial mass and surface rotational velocity by fitting Eq.~(\ref{eq_SNA_1}) 
to the data of the stellar model grids. The result is shown in Fig.~\ref{fig_coeff_c} on the left panel. The data is 
fitted well as a function of $v_\mathrm{ZAMS}$ using the polynomial function of second order 
\begin{eqnarray}
c = c_1 \cdot v_\mathrm{ZAMS}^2 + c_2 \cdot V_\mathrm{ZAMS} + c_3. \nonumber
\end{eqnarray}
The coefficients $c_1$, $c_2$ and $c_3$ are obtained for different initial masses to consider a mass dependence.

In the case of LMC models the dataset seen on the left panel in Fig.~\ref{fig_coeff_c} can be reproduced well by a linear function. 
Therefore parameter $c_1(M)$ is set to zero. This is not the case for SMC and MW were the polynomial function need to be used.

Figure \ref{fig_coeff_c} shows the mass dependent coefficients $c_2(M)$ and $c_3(M)$ which can be
fitted well using linear functions shown by the solid lines. Using  $c_1(M)$, $c_2(M)$ and $c_3(M)$, 
$c \left( M \left[ M_\odot \right], v_\mathrm{ZAMS} \left[ km/s \right] \right)$ can be calculated by 
\begin{equation}
c =  - \frac{M - 29.5}{54} + \left( M + 7.3 \right) \cdot \frac{v_\mathrm{ZAMS}}{10455}. 
\label{eq_parac_comp}
\end{equation}

\begin{figure}[htbp]
    \centering
        \includegraphics[width=8.5cm]{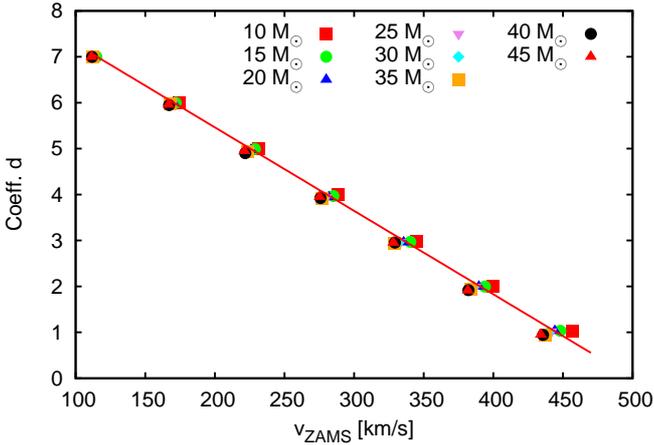}
    \caption{The parameter $d$ is shown as a function of initial surface rotational 
velocity for several different masses, obtained by fitting Eq.~(\ref{eq_SNA_1}) to the 
stellar evolution grid data (LMC). The line represents the best fit using a linear approximation. 
A small mass dependence can be seen, which will be neglected for simplification.}
    \label{fig_coeffd}
\end{figure}

Parameter $d$ is obtained for different masses and surface rotational velocities at the ZAMS. The resulting values are shown 
in Fig.~\ref{fig_coeffd}.  It can be seen, that only a weak mass dependence exists in the case of our data. For simplification, 
we neglect the mass dependence. Therefore a linear approximation is used to describe parameter $d$ as depicted in 
Fig.~\ref{fig_coeffd}, which can be described by
\begin{equation}
d = 9.1 - \frac{v_\mathrm{ZAMS}}{55}. 
\label{eq_parad_comp}
\end{equation}

\subsection{The main sequence lifetime and the surface rotational velocity}
\label{sec2_Cmslairv}
The fraction of the main sequence lifetime is used in Eq.~(\ref{eq_SNA_1}) to calculate the surface
nitrogen abundance. To be able to constrain the absolute age of the star it is necessary to know the 
time a star spends on the main sequence for given initial parameters.

\begin{figure*}[htbp]
    \centering
        \includegraphics[width=17cm]{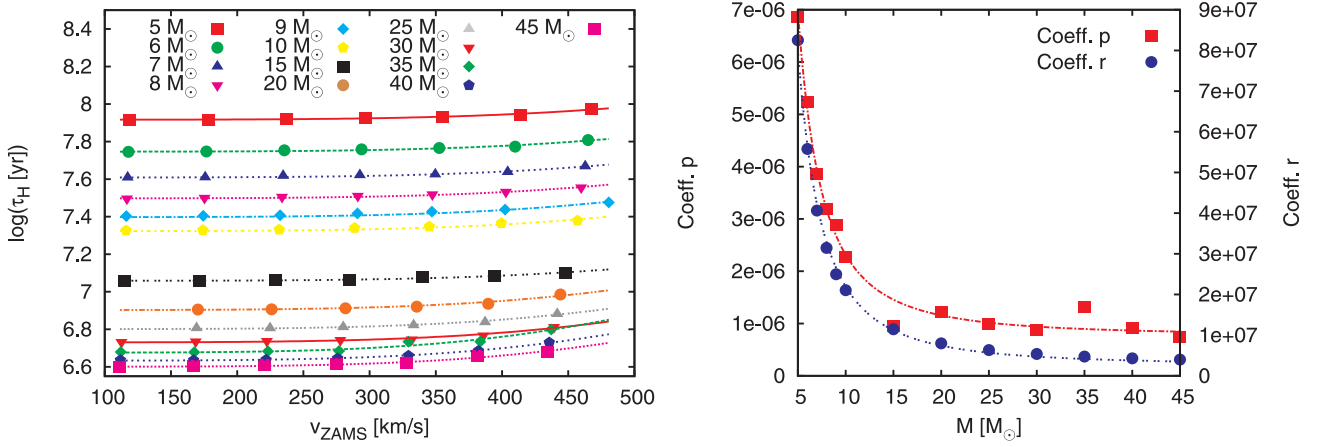}
    \caption{On the left panel the main sequence lifetime is shown as a function of the initial surface rotational velocity. 
Stellar evolution grid models for LMC metallicity are depicted for several different initial masses from 5 to 45\,$M_{\odot}$. 
The data can be fitted well using Eq.~(\ref{eq_tauMS}) as shown by the additional lines. On the right panel the two 
coefficients $p$ and $r$ are depicted as a function of mass. The dotted lines represent 
the best fit using Eq.~(\ref{eq_tauparameter}).}
    \label{fig_frac_MS_lifetime}
\end{figure*}

We therefore derive a formula for the main sequence lifetime for stars of a given initial mass and surface rotational 
velocity with MW, LMC and SMC metallicity. Figure~\ref{fig_frac_MS_lifetime} shows the main sequence lifetime as a 
function of the initial surface rotational velocity for the example 
of the LMC data for different initial masses. 
The higher the initial surface rotational velocity the higher the main sequence lifetime. Considering 
Fig.~\ref{fig_frac_MS_lifetime} it is noticeable that the main sequence lifetime is a strong function of the initial mass, 
but shows only a weak dependence on the initial surface rotational velocity. The data can be fitted well using 
\begin{equation}
\frac{\tau_\mathrm{H}}{\mathrm{yr}} =  p \cdot \left(v_\mathrm{ZAMS}\right)^{q} + r,
\label{eq_tauMS}
\end{equation}
taking mass dependent parameters $p, q$ and $r$ into account. For reproducing the data best, $q$ is chosen to be a linear 
function of mass, which in case of LMC and MW can be simplified to a constant value. $p$ and $r$ can be 
calculated depending on the initial mass. We therefore have 
\begin{eqnarray}
p &=& p_1 \cdot M^{p_2} + p_3, \nonumber \\
q &=& q_1 \cdot M + q_3, \nonumber \\
r &=& r_1 \cdot M^{r_2} + r_3. \nonumber
\label{eq_tauparameter}
\end{eqnarray}

By fitting the parameters $p$, $q$ and $r$, we gain the following formula to calculate the main sequence lifetime:
\begin{eqnarray}
\frac{\tau_\mathrm{H}}{\mathrm{yr}} &=&  \left( \frac{M^{-2} + 4.9804\cdot 10^{-3}}{6.4689\cdot 10^{3}} \right) \cdot \left(v_\mathrm{ZAMS} \right)^{4.57} \nonumber \\
&& + \frac{M^{-2} +1.3251 \cdot 10^{-3}}{5.1564\cdot 10^{-10}}.
\end{eqnarray}

\begin{figure*}[htbp]
    \centering
        \includegraphics[width=17cm]{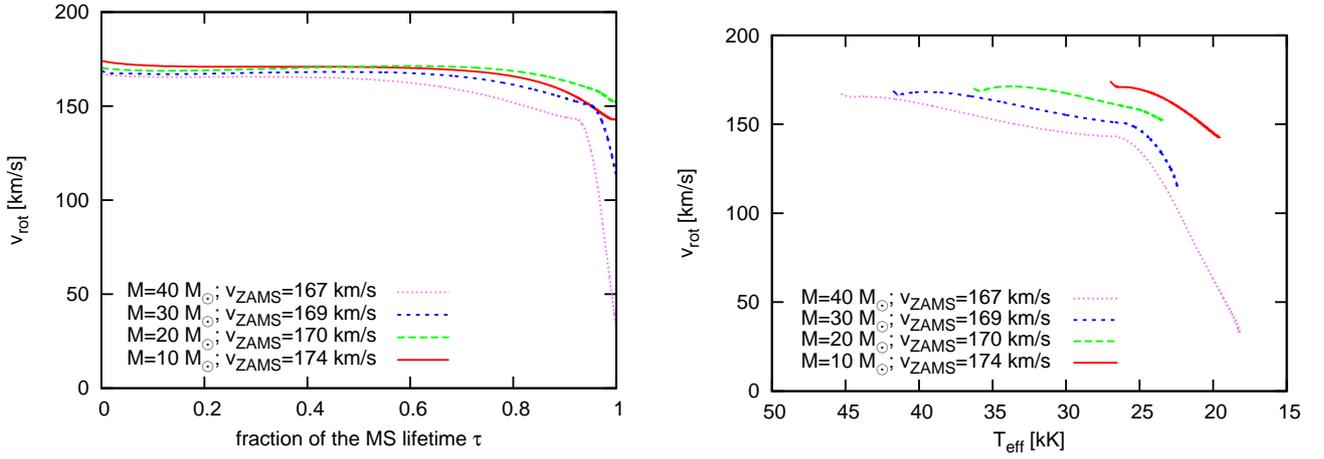}
    \caption{The evolution of the surface rotational velocity over time is shown on the left panel, while the surface 
rotational velocity as a function of the effective temperature is given on the right panel. Models of the stellar 
evolution grid in the case of LMC metallicity are used for different initial masses and surface rotational 
velocities of about 170\,km/s.}
    \label{fig_EVR}
\end{figure*}

In our stellar evolution models, mass loss is included as described in \citet{Vink_2010}. During the main sequence evolution the stars 
loses angular momentum due to mass loss. The value of the surface rotational velocity therefore changes as a function of time. 
Figure~\ref{fig_EVR} depicts the surface rotational velocity of a few exemplary models 
as functions of time (left panel) and effective temperature $T_\mathrm{eff}$ (right panel). 
An initial surface rotational velocity of about 170\,km/s is chosen.

During the evolution on the main sequence the surface rotational velocity of a star remains almost constant. At the end a sudden 
drop is visible for the highest considered masses, caused by an increase in mass loss and therefore a decrease of angular momentum 
when the effective temperature falls below 25\,000\,K, caused by the bi-stability braking \citep{Vink_2000,Vink_2001}. 
Stars with masses up to 30\,$M_\odot$ reach the end of the main sequence at temperatures higher than 20\,000\,K (see Fig.~\ref{fig_EVR}). 
They do not stay long enough on the main sequence at $T_\mathrm{eff} \le 25000\,$K to be significantly slowed down. In the case 
of stars more massive than 30\,$M_\odot$ the assumption is valid for $T_\mathrm{eff} \ge 25000\,$K.
For stars which have not been influenced noticeably by this effect it is reasonable to estimate the surface rotational velocity 
by the value at the ZAMS.

\subsection{Small Magellanic Cloud}
For the SMC, the parameters specifying Eq.~(\ref{eq_SNA_1}) are:
\begin{eqnarray}
a &=& 6.5, \nonumber \\
b &=& - \frac{M + 0.9}{1.4} + \left( M + 3.4 \right) \cdot \frac{v_\mathrm{ZAMS}}{122} - \left( M + 2.0 \right) \cdot \left( \frac{v_\mathrm{ZAMS}}{262} \right)^2, \nonumber \\
c &=& - \frac{M - 71}{67} + \left( M - 40 \right) \cdot \frac{v_\mathrm{ZAMS}}{18457} + \left( M + 213 \right) \cdot \left( \frac{v_\mathrm{ZAMS}}{6294} \right)^2,\nonumber \\
d &=& 9.0 - \frac{v_\mathrm{ZAMS}}{58}, \nonumber
\end{eqnarray}
\begin{eqnarray}
\frac{\tau_\mathrm{H}}{\mathrm{yr}} &=&  \left( \frac{M^{-5.85}}{8.5881\cdot 10^{-4}} \right) \cdot \left(v_\mathrm{ZAMS}\right)^{0.1121 \cdot M + 2.4187} \nonumber \\
&& + \frac{M^{-2.065} + 1.6507\cdot 10^{-3}}{5.0320\cdot 10^{-10}}.\nonumber
\end{eqnarray}

The formula to calculate the main sequence lifetime for SMC stars is valid for all stars analyzed, but not in the complete range 
given in Table~\ref{tab_valcond}. Table~\ref{tab_valcondSMCtau} summarizes the mass and velocity ranges valid for this specific formula.

\begin{table}[h]
\caption{Range of initial stellar parameters for which the main sequence lifetime calculation for SMC stars is validated.} 
\label{tab_valcondSMCtau}  
\centering                                  
\begin{tabular}{c c}  
\hline\hline
$v_\mathrm{ZAMS}\,$ &  $M$\\
 km/s & $M_\odot$ \\
\hline
0--50 & 5--50 \\
50--100 & 5--30 \\
100--225 & 5--21 \\
225--350 & 5--15 \\
\hline
\end{tabular}
\end{table}

\subsection{Milky Way}
For the MW, the parameters specifying Eq.~(\ref{eq_SNA_1}) are:
\begin{eqnarray}
a &=& 7.64, \nonumber \\
b &=& - \frac{M + 20.3}{4.0} + \left( M + 10.3 \right) \cdot \frac{v_\mathrm{ZAMS}}{204} - \left( M + 9.9 \right) \cdot \left( \frac{v_\mathrm{ZAMS}}{338} \right)^2, \nonumber \\
c &=& - \frac{M - 86}{69} + \left( M - 137 \right) \cdot \frac{v_\mathrm{ZAMS}}{22561} + \left( M + 240 \right) \cdot \left( \frac{v_\mathrm{ZAMS}}{4737} \right)^2,  \nonumber \\
d &=& 9.1 - \frac{v_\mathrm{ZAMS}}{53}, \nonumber
\end{eqnarray}
\begin{eqnarray}
\frac{\tau_\mathrm{H}}{\mathrm{yr}} &=&  \left( \frac{M^{-2.05} + 2.3107\cdot 10^{-3}}{1.6504\cdot 10^{-1}} \right) \cdot \left(v_\mathrm{ZAMS}\right)^{2.89} \nonumber \\
&& + \frac{M^{-2.35} + 1.1714\cdot 10^{-3}}{2.6032 \cdot 10^{-10}}.\nonumber 
\end{eqnarray}

\section{Dataset}
\label{app2}

\begin{table*}
\caption{Properties of the analyzed LMC stars.}          
\label{table_dataLMC}  
\centering                                  
\begin{tabular}{l r r r r r r r r r r c r}
\hline
\hline                      
Star & Number & $v\sin \left(i \right)$ & $T_{\mathrm{eff}}$ & $M_{\mathrm{evol}}$  & $\mathcal N$ & $\tau_\mathrm{H}$ & $\sin \left( i_{\mathrm{N}} \right)$ & $\sin \left( i_{\mathrm{iso}} \right)$ & $t_{\mathrm{N}}$ & $t_{\mathrm{isochrone}}$ & binary & class\\  
    & &  km/s                &  K      &  $M_\odot$ &  dex      &  Myr      &                    &          &  Myr          &  Myr          &        & \\
\hline                                
\object{N 11--34}  & L1  & 203 & 25500 & 20 & 7.12      & 7.5  & 1.00 & --   & $2.7^{+1.0}_{-1.5}$   & 7.8  & yes & 1 \\
\object{N 11--37}  & L2  & 100 & 28100 & 23 & $\le$7.17 & 6.2  & 0.76 & $\ge$0.76 & $6.2$            & 6.8  & yes & 2 \\
\object{N 11--46}  & L3  & 205 & 33500 & 28 & $\le$7.71 & 5.1  & 0.88 & $\ge$0.88 & $5.1_{-2.5}$      & 4.8  & yes & 2 \\
\object{N 11--77}  & L4  & 117 & 21500 & 12 & 6.96      & 16.0 & 0.90 & 0.89 & $16.0_{-16.0}$    & 15.4 & yes & 2 \\ 
\object{N 11--88}  & L5  & 240 & 24150 & 14 & $\le$6.96 & 12.6 & 1.00 & --   & $2.6^{+1.9}_{-2.6}$   & 12.6 & --  & 1 \\
&&&&&&&&&&&&\\
\object{N 11--89}  & L6  & 117 & 21700 & 12 & 6.91      & 16.0 & 1.00 & --   & $14.6^{+1.4}_{-14.6}$ & 16.4 & yes & 1 \\
\object{N 11--95}  & L7  & 267 & 26800 & 15 & 7.46      & 11.4 & 1.00 & --   & $4.9^{+2.4}_{-1.3}$   & 10.4 & --  & 1 \\
\object{N 11--102} & L8  & 218 & 31000 & 18 & $\le$7.58 & 8.6  & 1.00 & --   & $6.6^{+2.0}_{-2.2}$   & 7.0  & --  & 1 \\
\object{N 11--104} & L9  & 153 & 25700 & 14 & $\le$7.20 & 12.5 & 0.99 & $\ge$0.97 & $12.5_{-4.2}$     & 11.8 & --  & 2 \\
\object{N 11--111} & L10 & 101 & 22950 & 12 & 6.93      & 16.0 & 0.83 & 0.81 & $16.0_{-16.0}$    & 15.4 & --  & 2 \\
&&&&&&&&&&&&\\
\object{N 11--114} & L11 & 299 & 32500 & 19 & $\le$7.92 & 8.2  & 0.91 & $\ge$0.90 & $8.2_{-4.7}$      & 5.8  & --  & 2 \\
\object{N 11--116} & L12 & 160 & 21700 & 11 & 6.98      & 18.6 & 1.00 & --   & $12.4^{+6.2}_{-12.4}$ & 18.4 & --  & 1 \\
\object{N 11--117} & L13 & 247 & 26800 & 13 & $\le$7.06 & 14.2 & 1.00 & --   & $4.1^{+1.9}_{-4.1}$   & 11.2 & --  & 1 \\
\object{N 11--118} & L14 & 150 & 25700 & 13 & 7.39      & 14.1 & 0.83 & 0.79 & $14.1$           & 12.6 & yes & 2 \\
\object{N 11--120} & L15 & 207 & 31500 & 17 & $\le$7.56 & 9.3  & 1.00 & 0.91 & $8.2^{+1.1}_{-2.8}$   & 6.4  & --  & 1 \\
&&&&&&&&&&&&\\
\object{N 11--121} & L16 & 265 & 27000 & 14 & $\le$7.48 & 12.6 & 1.00 & --   & $6.1^{+3.4}_{-1.6}$   & 11.2 & --  & 1 \\
\object{N 11--122} & L17 & 173 & 33000 & 18 & $\le$7.72 & 8.6  & 0.68 & $\ge$0.61 & $8.6$            & 4.6  & --  & 2 \\
\object{NGC 2004--20}  & L18 & 145 & 22950 & 18 & 7.46      & 8.6  & 0.83 & 0.83 & $8.6$            & 9.6  & yes & 2 \\
\object{NGC 2004--30}  & L19 & 123 & 29000 & 19 & 7.43      & 7.9  & 0.71 & 0.71 & $7.9$            & 7.6  & yes & 2 \\
\object{NGC 2004--47}  & L20 & 133 & 21700 & 12 & 7.03      & 16.0 & 0.94 & 0.94 & $16.0_{-16.0}$    & 16.4 & yes & 2 \\ 
&&&&&&&&&&&&\\
\object{NGC 2004--50}  & L21 & 109 & 20450 & 11 & 7.10      & 18.6 & 0.71 & 0.68 & $18.6_{-18.6}$    & 17.0 & yes & 2 \\
\object{NGC 2004--54}  & L22 & 114 & 21700 & 12 & 6.97      & 16.0 & 0.86 & 0.86 & $16.0_{-16.0}$    & 16.4 & yes & 2 \\
\object{NGC 2004--62}  & L23 & 106 & 31750 & 17 & $\le$7.33 & 9.3  & 0.67 & $\ge$0.61 & $9.3$            & 7.6  & --  & 2 \\
\object{NGC 2004--63}  & L24 & 107 & 21700 & 11 & 6.97      & 18.6 & 0.79 & 0.79 & $18.6_{-18.6}$    & 18.0 & --  & 2 \\
\object{NGC 2004--65}  & L25 & 223 & 20450 & 11 & $\le$7.19 & 18.7 & 1.00 & --   & $9.7^{+3.7}_{-3.6}$   & 19.8 & --  & 1 \\
&&&&&&&&&&&&\\
\object{NGC 2004--66}  & L26 & 238 & 25700 & 13 & 7.98      & 14.2 & 0.60 & 0.59 & $14.2$           & 12.0 & --  & 2 \\
\object{NGC 2004--69}  & L27 & 178 & 28000 & 15 & 7.84      & 11.2 & 0.57 & 0.56 & $11.2$           & 9.8  & --  & 2 \\
\object{NGC 2004--74}  & L28 & 130 & 27375 & 14 & $\le$7.42 & 12.5 & 0.72 & $\ge$0.67 & $12.5$           & 10.6 & yes & 2 \\
\object{NGC 2004--75}  & L29 & 116 & 21700 & 11 & 7.06      & 18.6 & 0.78 & 0.77 & $18.4_{-18.6}$    & 18.0 & --  & 2 \\
\object{NGC 2004--77}  & L30 & 215 & 29500 & 16 & 7.65      & 10.2 & 0.93 & 0.90 & $10.2_{-3.6}$     & 8.4  & --  & 2 \\
&&&&&&&&&&&&\\
\object{NGC 2004--79}  & L31 & 165 & 21700 & 11 & 7.60      & 18.6 & 0.70 & 0.70 & $18.6$           & 18.4 & yes & 2 \\
\object{NGC 2004--81}  & L32 & 105 & 26800 & 13 & 7.14      & 14.0 & 0.70 & 0.63 & $14.0$           & 11.2 & --  & 2 \\
\object{NGC 2004--82}  & L33 & 161 & 25700 & 13 & 7.47      & 14.1 & 0.83 & 0.79 & $14.1$           & 12.6 & --  & 2 \\
\object{NGC 2004--85}  & L34 & 150 & 20450 & 10 & 7.25      & 22.0 & 0.85 & 0.84 & $22.0$           & 21.4 & --  & 2 \\
\object{NGC 2004--88}  & L35 & 200 & 20450 & 10 & $\le$7.18 & 22.0 & 1.00 & --   & $15.1^{+5.7}_{-5.7}$  & 21.6 & yes & 1 \\
&&&&&&&&&&&&\\
\object{NGC 2004--95}  & L36 & 138 & 25700 & 13 & 7.13      & 14.1 & 0.92 & 0.89 & $14.1_{-4.8}$     & 12.8 & --  & 2 \\
\object{NGC 2004--99}  & L37 & 119 & 21700 & 10 & 7.06      & 22.0 & 0.78 & 0.74 & $22.0_{-22.0}$    & 19.6 & --  & 2 \\
\object{NGC 2004--100} & L38 & 323 & 26800 & 13 & $\le$7.29 & 14.5 & 1.00 & --   & $3.6^{+1.3}_{-1.2}$   & 11.6 & --  & 1 \\
\object{NGC 2004--101} & L39 & 131 & 21700 & 10 & 7.59      & 22.0 & 0.55 & 0.53 & $22.0$           & 19.6 & --  & 2 \\
\object{NGC 2004--104} & L40 & 274 & 25700 & 12 & $\le$7.28 & 16.3 & 1.00 & --   & $6.0^{+2.2}_{-1.8}$   & 13.2 & --  & 1 \\
&&&&&&&&&&&&\\
\object{NGC 2004--105} & L41 & 235 & 25700 & 12 & $\le$7.11 & 16.2 & 1.00 & --   & $6.1^{+2.5}_{-3.7}$   & 13.2 & --  & 1 \\
\object{NGC 2004--107} & L42 & 146 & 28500 & 14 & $\le$7.12 & 12.5 & 1.00 & 0.88 & $12.3^{+0.2}_{-6.2}$  & 9.4  & yes & 1 \\
\object{NGC 2004--110} & L43 & 121 & 21700 & 10 & 6.95      & 22.0 & 0.91 & 0.87 & $22.0_{-22.0}$    & 19.4 & --  & 2 \\
\object{NGC 2004--113} & L44 & 307 & 20450 & 9  & $\le$7.29 & 27.1 & 1.00 & --   & $9.5^{+3.6}_{-3.1}$   & 22.2 & --  & 1 \\
\hline                                          
\end{tabular}
\tablefoot{From left to right: catalogue number (Evans et al. 2005), reference number used in this paper, 
projected surface rotational velocity, effective temperature, evolutionary mass and surface nitrogen abundance
(all from Hunter et al. 2008b, 2009), main sequence lifetime, nitrogen inclination angle, inclination angle derived using 
the nitrogen abundance and the isochrone age, maximum age based on the surface nitrogen abundance, isochrone age, 
identification of binarity and class. 
For Class~2 stars, adopting an error in the nitrogen abundance does not lead to an increased upper age limit,
and in many cases also not to an decreased lower age limit.}
\end{table*}

\begin{table*}
\caption{Properties of the analyzed SMC stars.}
\label{table_dataSMC}  
\centering                                  
\begin{tabular}{l r r r r r r r r r r c r}
\hline
\hline
Star & Number & $v\sin \left(i \right)$ & $T_{\mathrm{eff}}$ & $M_{\mathrm{evol}}$  & $\mathcal N$ & $\tau_\mathrm{H}$ & $\sin \left( i_{\mathrm{N}} \right)$ & $\sin \left( i_{\mathrm{iso}} \right)$ & $t_{\mathrm{N}}$ & $t_{\mathrm{isochrone}}$ & binary & class\\  
    & &  km/s                &  K      &  $M_\odot$ &  dex      &  Myr      &           &                   &  Myr          &  Myr          &        & \\
\hline                                
\object{NGC 346--27} & S1  & 220 & 31000 & 18 & $\le$7.71 & 9.7  & 0.57 & $\ge$0.57 & $9.7$          & 7.6  & --  & 2 \\
\object{NGC 346--32} & S2  & 125 & 29000 & 17 & $\le$6.88 & 9.1  & 0.85 & $\ge$0.85 & $9.1$          & 9.2  & yes & 2 \\
\object{NGC 346--53}  & S3  & 170 & 29500 & 15 & $\le$7.44 & 10.9 & 0.57 & $\ge$0.57 & $10.9$         & 9.4  & yes & 2 \\
\object{NGC 346--55}  & S4  & 130 & 29500 & 15 & $\le$7.25 & 10.9 & 0.64 & $\ge$0.58 & $10.9$         & 9.4  & --  & 2 \\
\object{NGC 346--58}  & S5  & 180 & 29500 & 14 & $\le$7.56 & 12.0 & 0.51 & $\ge$0.51 & $12.0$         & 9.6  & yes & 2 \\
&&&&&&&&&&&&\\
\object{NGC 346--70}  & S6  & 109 & 30500 & 15 & $\le$7.50 & 11.0 & 0.34 & $\ge$0.33 & $11.0$         & 8.2  & --  & 2 \\
\object{NGC 346--79}  & S7  & 293 & 29500 & 14 & $\le$7.88 & 13.4 & 0.65 & $\ge$0.64 & $13.4$         & 9.4  & --  & 2 \\
\object{NGC 346--80}  & S8  & 216 & 27300 & 12 & $\le$7.85 & 15.4 & 0.48 & $\ge$0.48 & $15.4$         & 12.8 & --  & 2 \\
\object{NGC 346--81}  & S9  & 255 & 21200 & 9  & $\le$7.52 & 25.1 & 0.73 & $\ge$0.73 & $25.1$         & 24.4 & --  & 2 \\
\object{NGC 346--82}  & S10 & 168 & 21200 & 9  & $\le$7.55 & 24.7 & 0.46 & $\ge$0.46 & $24.7$         & 24.4 & yes & 2 \\
&&&&&&&&&&&&\\
\object{NGC 346--83}  & S11 & 207 & 27300 & 12 & $\le$7.73 & 15.3 & 0.50 & $\ge$0.50 & $15.3$         & 12.8 & yes & 2 \\
\object{NGC 346--84}  & S12 & 105 & 27300 & 12 & $\le$7.06 & 15.0 & 0.59 & $\ge$0.55 & $15.2$         & 12.8 & --  & 2 \\
\object{NGC 346--92}  & S13 & 234 & 27300 & 12 & $\le$7.52 & 15.5 & 0.69 & $\ge$0.68 & $15.5$         & 12.8 & --  & 2 \\
\object{NGC 346--100} & S14 & 183 & 26100 & 11 & $\le$7.41 & 17.5 & 0.63 & $\ge$0.61 & $17.5$         & 15.0 & --  & 2 \\
\object{NGC 346--106} & S15 & 142 & 27500 & 12 & $\le$7.51 & 15.1 & 0.42 & $\ge$0.42 & $15.1$         & 12.4 & yes & 2 \\
&&&&&&&&&&&&\\
\object{NGC 346--108} & S16 & 167 & 26100 & 11 & $\le$7.42 & 17.4 & 0.56 & $\ge$0.55 & $17.4$         & 15.0 & --  & 2 \\
\object{NGC 346--109} & S17 & 123 & 26100 & 11 & $\le$7.09 & 17.4 & 0.67 & $\ge$0.62 & $17.4$         & 15.0 & --  & 2 \\
\object{NGC 346--114} & S18 & 287 & 27300 & 12 & $\le$7.80 & 16.0 & 0.66 & $\ge$0.65 & $16.0$         & 12.6 & --  & 2 \\
\object{NGC 330--21}  & S19 & 204 & 30500 & 21 & $\le$7.64 & 9.2  & 0.57 & $\ge$0.57 & $9.2$          & 7.0  & --  & 2 \\
\object{NGC 330--38}  & S20 & 150 & 27300 & 14 & $\le$7.16 & 11.9 & 0.81 & $\ge$0.79 & $11.9$         & 11.2 & --  & 2 \\
&&&&&&&&&&&&\\
\object{NGC 330--39}  & S21 & 120 & 33000 & 18 & $\le$7.61 & 8.4  & 0.34 & $\ge$0.34 & $8.4$          & 6.4  & --  & 2 \\
\object{NGC 330--40}  & S22 & 106 & 21200 & 11 & $\le$7.12 & 17.4 & 0.56 & $\ge$0.56 & $17.4$         & 20.6 & --  & 2 \\
\object{NGC 330--41}  & S23 & 127 & 32000 & 17 & 7.73      & 9.1  & 0.32 & 0.32 & $9.1$          & 7.2  & --  & 2 \\
\object{NGC 330--45}  & S24 & 133 & 18450 & 9  & $\le$7.28 & 24.6 & 0.56 & $\ge$0.56 & $24.6$         & 29.0 & yes & 2 \\
\object{NGC 330--51}  & S25 & 273 & 26100 & 12 & $\le$7.20 & 15.8 & 1.00 & --   & $8.7^{+7.1}_{-2.2}$ & 14.2 & --  & 1 \\
&&&&&&&&&&&&\\
\object{NGC 330--56}  & S26 & 108 & 21200 & 9  & $\le$6.93 & 24.6 & 0.63 & $\ge$0.62 & $24.6$         & 24.0 & --  & 2 \\
\object{NGC 330--57}  & S27 & 104 & 29000 & 13 & $\le$7.48 & 13.3 & 0.33 & $\ge$0.32 & $13.3$         & 10.2 & --  & 2 \\
\object{NGC 330--59}  & S28 & 123 & 19000 & 8  & 7.27      & 30.5 & 0.51 & 0.51 & $30.5$         & 30.8 & --  & 2 \\
\object{NGC 330--66}  & S29 & 126 & 18500 & 8  & $\le$7.77 & 30.5 & 0.28 & $\ge$0.28 & $30.5$         & 34.6 & --  & 2 \\
\object{NGC 330--69}  & S30 & 193 & 19000 & 8  & $\le$7.91 & 30.6 & 0.32 & $\ge$0.32 & $30.6$         & 33.2 & --  & 2 \\
&&&&&&&&&&&&\\
\object{NGC 330--79}  & S31 & 146 & 19500 & 8  & $\le$7.69 & 30.5 & 0.35 & $\ge$0.35 & $30.5$         & 32.6 & yes & 2\\
\object{NGC 330--83}  & S32 & 140 & 18000 & 7  & $\le$7.66 & 39.1 & 0.34 & $\ge$0.34 & $39.1$         & 39.8 & --  & 2 \\
\object{NGC 330--97}  & S33 & 154 & 27300 & 11 & 7.60      & 17.4 & 0.41 & 0.40 & $17.4$         & 12.0 & --  & 2 \\
\object{NGC 330--110} & S34 & 110 & 21200 & 8  & 7.23      & 30.4 & 0.48 & 0.48 & $30.4$         & 30.0 & --  & 2 \\
\object{NGC 330--120} & S35 & 137 & 18500 & 6  & $\le$7.85 & 52.5 & 0.28 & $\ge$0.28 & $52.5$         & 46.8 & --  & 2 \\
\hline                                          
\end{tabular}
\tablefoot{From left to right: catalogue number (Evans et al. 2005), reference number used in this paper, 
projected surface rotational velocity, effective temperature, evolutionary mass and surface nitrogen abundance
(all from Hunter et al. 2008b, 2009), main sequence lifetime, nitrogen inclination angle, inclination angle derived using 
the nitrogen abundance and the isochrone age, maximum age based on the surface nitrogen abundance, isochrone age, 
identification of binarity and class. 
For Class~2 stars, adopting an error in the nitrogen abundance does not lead to an increased upper age limit,
and in many cases also not to an decreased lower age limit.}
\end{table*}

\end{appendix}

\bibliographystyle{aa}
\bibliography{literatur}

\listofobjects
\end{document}